\documentclass[12pt]{iopart}
\usepackage{graphicx}
\newcommand{\<}{\left\langle}
\renewcommand{\>}{\right\rangle}
\renewcommand{\{}{\left\lbrace}
\renewcommand{\}}{\right\rbrace}

\newcommand{\bea}{\begin{eqnarray}}
\newcommand{\eea}{\end{eqnarray}}
\newcommand{\pih}{{\hat \pi}}

\newcommand{\vS}{{\mbox{\boldmath{$S$}}}}
\newcommand{\vsigma}{{\mbox{{\boldmath{$\sigma$}}}}}
\newcommand{\vb}{{\mbox{\boldmath{$b$}}}}

\newcommand{\vh}{{\mbox{\boldmath{$h$}}}}
\newcommand{\vx}{{\mbox{\boldmath{$x$}}}}
\newcommand{\vv}{{\mbox{\boldmath{$v$}}}}
\newcommand{\matJ}{{\mbox{\boldmath{$J$}}}}
\newcommand{\matA}{{\mbox{\boldmath{$A$}}}}
\newcommand{\matx}{{\mbox{\boldmath{$x$}}}}
\newcommand{\qs}{{q_{\<\>}}}

\newcommand{\cut}[1]{\hspace{-1ex}}
\newcommand{\alal}{{\<\alpha_1 \alpha_2\>}}
\newcommand{\alalal}{{\<\alpha_1 \alpha_2 \alpha_3\>}}
\newcommand{\alalalal}{{\<\alpha_1 \alpha_2 \alpha_3 \alpha_4\>}}
\newcommand{\ij}{{\< i j \>}}

\newcommand{\qalal}{{q_\alal}}
\newcommand{\qhalal}{{ {\hat q}_\alal}}

\newcommand{\qal}{{q_\alpha}}
\newcommand{\tal}{{t_\alpha}}
\newcommand{\qbaral}{{\bar {q}}_\alpha}
\newcommand{\qbaro}{{\bar q}_1}

\newcommand{\Qalal}{Q_{\alpha_1 \alpha_2}}

\newcommand{\qhbaral}{{\hat \qbaral}}
\newcommand{\Jt}{{\tilde J}}
\newcommand{\Jo}{{J_0}}
\newcommand{\bbar}{{\bar b}}
\newcommand{\SP}{{\left(\< \exp \{\beta x \sum_\alpha \sigma_1^{\alpha}
\sigma_2^{\alpha} \}\>_{\phi(x)} -1\right)}}
\newcommand{\atanh}{{\tt atanh}}

\newcommand{\Avb}[1]{\left\langle #1 \right\rangle_b}
\newcommand{\mT}{\mathcal{T}}
\newcommand{\mI}{\mathcal{I}}
\newcommand{\mX}{\mathcal{X}}

\newcommand{\mW}{\mathcal{W}}
\newcommand{\mZ}{\mathcal{Z}}
\newcommand{\mA}{\mathcal{A}}
\newcommand{\mB}{\mathcal{B}}
\newcommand{\mC}{\mathcal{C}}
\newcommand{\mD}{\mathcal{D}}
\newcommand{\mP}{\mathcal{P}}
\newcommand{\mQ}{\mathcal{Q}}
\newcommand{\mR}{\mathcal{R}}
\newcommand{\fatI}{\mathbf{I}}

\newcommand{\Qalalal}{Q_{\alpha_1 \alpha_2 \alpha_3}}
\newcommand{\Qalalalal}{Q_{\alpha_1 \alpha_2 \alpha_3 \alpha_4}}
\newcommand{\Qoalal}{{\bar Q}_{\<\alpha_1 \alpha_2\>}}
\newcommand{\Qoalalal}{Q_{\<\alpha_1 \alpha_2 \alpha_3\>}}
\newcommand{\Qoalalalal}{Q_{\<\alpha_1 \alpha_2 \alpha_3 \alpha_4\>}}

\begin{document}

\title[On composite systems]{On composite systems of dilute and dense couplings}
\author{J. R. Raymond and D. Saad}
\address{Aston University, Neural Computing Research Group (NCRG),
Birmingham, B4 7ET, UK}
\ead{jack.raymond@physics.org}
\begin{abstract}
Composite systems, where couplings are of two types, a combination
of strong dilute and weak dense couplings of Ising spins, are
examined through the replica method. The dilute and dense parts
are considered to have independent canonical disordered or uniform
bond distributions; mixing the models by variation of a parameter
$\gamma$ alongside inverse temperature $\beta$ we analyse the
respective thermodynamic solutions. We describe the variation in
high temperature transitions as mixing occurs; in the vicinity of
these transitions we exactly analyse the competing effects of the
dense and sparse models. By using the replica symmetric ansatz and
population dynamics we described the low temperature behaviour of
mixed systems.
\end{abstract}

\section{Introduction}

Understanding phenomena arising in many body systems through mean
field analysis of simple models has provided insight into many
problems in physics, theoretical computer science,
telecommunication, biology and
elsewhere~\cite{Mezard:SGT,Nishimori:SP}.
Statistical mechanics describes aspects of macroscopic behaviour
in interacting systems of many elements, and methods originating
in the study of spin-glasses (SG) have been extended to explore
many interesting model disordered systems. These statistical
descriptors of behaviour often prove to be a sufficient descriptor
of all interesting bulk behaviour; in some applications, such as
channel coding and theoretical computer science, they also provide
benchmarks even where {\it the large system} is relatively small
and the randomness assumed does not quite match the true system
conditions. Moreover, methods developed within the statistical
physics community gave rise to the development of efficient
inference algorithms widely used in telecommunication,
probabilistic modelling and theoretical computer science.

Many of these models
consider range-free interactions of a single topological type,
most commonly nearest neighbour interactions in finite
dimensional, fully connected, or sparse random graphs. Even in
systems not conforming strongly to the particular topology, insight
into many phenomena can be obtained and the formulations can
allow for exact analysis. We propose that certain systems may be
well described by a combination of two canonical topologies --
such a property may be phenomenological or could be a deliberately
engineered feature of a system.

Amongst the best understood topologies are those amenable to mean
field theory models that are {\it infinite dimensional}. The
canonical mean field model is that of a fully connected graph.
This model may be analysed exactly for the cases of uniform
interactions and certain random ensembles, most famously the
celebrated Sherrington Kirkpatrick (SK)
model~\cite{Sherrington:SM}. Simplification of the analysis in the
disordered case is often possible through noting the ability to
describe many combinations of interactions by a Gaussian due to
the central limit theorem. The sparse mean field model has each
spin coupled only to a small number of other spins. This creates a
topology that is also described as infinite dimensional though the
model is perceived as more representative of most complex systems
due to some notion of a neighbourhood being maintained. Analysis
is simplified by utilising the asymptotic cycle free property
(Bethe approximation) of certain random graph ensembles. Models
which do not allow use of the Bethe approximation or central limit
theorem are in general difficult to analyse.

The motivation for studying this system is that it is amongst the
simplest composite models involving a combination of dilute and
dense couplings. We anticipate the understanding of these systems
to be important in engineering applications. With miniaturisation
of technology, for example computer chips, the paradigm by which
the interaction amongst components can be strictly controlled may
be invalid. Modelling the effect of non-engineered couplings by
independent noise may be invalid in some scenarios, as it is
possible that these additional interactions form a strongly
correlated network allowing for example non-trivial phase
transitions or metastable features. There is also the possibility
that systems may be engineered deliberately with a combination of
dilute strong, and weak dense interactions either to make them
more robust or to exploit specific properties of the composite
system. In multi-user channel coding, for instance, this may make
the communication process more robust against different types of
noise, de-synchronisation or malicious attacks. Only recently, a
special case of the system studied here was suggested as a model
for studying the resilience of networks against
attacks~\cite{Hase}.

The paper is organised as follows. In section~\ref{sec:model} we
introduce the model studied and basic definitions, followed by an
analysis section~\ref{sec:analysis} that explains briefly the
derivation based on the replica method.  In
sections~\ref{sec:highT} and~\ref{sec:lowT} we investigate the
high and low temperature solutions, respectively, followed by a
discussion of the numerical solutions obtained via population
dynamics section~\ref{sec:results}. Finally, we will present our
conclusions in section~\ref{sec:conclusions}.

\section{The model}
\label{sec:model}

We present analysis for an ensemble of disordered systems of $N$
Ising spins ${s_i=\pm 1}$. The couplings consists of a strong part
which is non-zero on only a fraction $\rho N/2$ of the possible
links, and a weak part which is present on all links, this defines
the Hamiltonian consisting of sparse and dense parts
\begin{equation}
 {\cal H}(\vS) = {\cal H}_S (\vS) + {\cal H}_D (\vS) + \sum_i h_i S_i
\end{equation}
along with an external field. The dense and sparse parts are taken
to be
\begin{eqnarray}
 {\cal H}_D(\vS)  &=&  - \sum_{\langle i j \rangle} b_i b_j
 J^D_{\langle i j \rangle} \sigma_i \sigma_j,  \nonumber \ , \\
{\cal H}_S(\vS) &=& - \sum_{\langle i j \rangle} b^S_i b^S_j
A_{\langle i j \rangle} J^S_{\langle i j \rangle} \sigma_i
\sigma_j \; ,
\end{eqnarray}
where $\<\>$ indicates the set of ordered (distinct) indices
throughout the paper. Components of the sparse connectivity matrix
${\bf A}$ take the value 1 if a link exists between the
corresponding sites and zero otherwise; the coupling matrix ${\bf
J}$ takes random values from a given distribution and $\vb$
determines an {\it artificial} disorder (alignment) in the
coupling strengths, analogous to the Mattis model~\cite{Mattis}.
For our analysis we can take $\vb^S=1$ since only the relative
alignment is influential in determining system properties.

\subsubsection{Mixing of models}
In order to investigate the combination of these subsystems we
propose the introduction of two parameters, an inverse temperature
$\beta$, and a mixing parameter $\gamma \in (0,1)$. The mixing
parameter acts so that the couplings in the sparse part increase
monotonically from zero with $\gamma$ to their full value,
conversely the couplings of the dense part decrease monotonically
to zero. There is some flexibility in how this might be applied,
for example if the couplings decrease/increase linearly with
$\gamma$ one may write the Hamiltonian as
\begin{equation}
 {\cal H}(\vS) = \gamma {\cal H}_S (\vS) + (1-\gamma){\cal H}_D (\vS)
 + \sum_i h_i S_i \; . \label{Hamgam}
\end{equation}
This is the simplest composition method we may use but alternative
rescalings of the couplings may also be sensible. Unattractive
features include that the ratio of variance to mean ($\Jo/\Jt$) in
coupling strengths is not maintained as $\gamma$ varies.

The mixing of models by rescaling of the couplings is not a unique
way to consider the composition of two such systems. A sensible
alternative valid in some composite systems would be a doping one,
keeping the dense part constant and gradually introducing
additional sparse bonds (variation of $\matA$)~\cite{Hase}.

\subsubsection{Definition of the ensemble}

We consider different ensembles determined by a set of parameters
${\cal I}=\{J_0,{\tilde J},\rho,\phi,h^S,h^D,\chi,\bar{b}\}$
defined as follows. The model consists of independently
generated dense and sparse subsystems. Dense Couplings are sampled
independently for each link from a distribution of mean $J_0 b_i
b_j/N$ and variance ${\tilde J}/N$ and non-divergent higher
moments (the scaling in $N$ is standard~\cite{Sherrington:SM}).
The Mattis model-like part $\vb$ describes some non-trivial
orientation of the spins with $b_i=\pm 1$ sampled independently at
each site according to the mean $\bbar$. For the sparse part we
have that the connectivity matrix, ${\bf A}$, is a sample from an
Erd\"{o}s-Renyi random graph ensemble~\cite{ErR70} of mean
connectivity $\rho$ with couplings sampled independently for each
link from a distribution on the real line, $\phi$, with
non-divergent moments, and vanishing measure on zero. The external fields
are sampled for each site from a distribution of mean
$h^S + b_i h^D$, and variance $\chi^2$, the parameters $h_S$,$h_D$ and $\chi^2$ are conjugate
variables in the free energy to the order parameters for sparse and 
dense aligned ferromagnetic moments, and the spin glass moment.

The model contains a wide range of parameters which we believe
are sufficient to describe the mixing of many canonical Ising spin
mean field models. In the case of $\gamma\rightarrow0$ the model
reduces to the SK model~\cite{Sherrington:SM} (up to artificial
disorder), whereas at $\gamma \rightarrow 1$ the model is Viana
Bray (VB)~\cite{Viana:PD}. By tuning parameters one can also find
the ferromagnetic, antiferromagnetic and Mattis
models~\cite{Mattis}, the relevant orientation in mixing ($\bbar$)
proves important in determining system properties. Small
perturbations of the SK and VB models by random Hamiltonians is a
subject much studied, especially in the context of temperature
variation and stochastic
stability~\cite{Barra:SS,Parisi:stability,Talagrand}. We
understand that the set of perturbations represented by variation
of an infinitesimal variation of $\gamma$ from $0$ or $1$ probably
falls into the classes which are incapable of changing the
structure of thermodynamic states - provided we break any
interaction symmetries by addition of small external fields, and
so we expect no transitions in these limits, which is both an
observed and intuitive assumption.

\section{Replica method and exact analysis}
\label{sec:analysis}
\subsection{Self-averaged free energy calculation}

The analysis is rather standard and is carried out by the replica
method, we note only important key steps and definitions in this
section. Details of the derivation are found
in~\ref{app.RC}. Throughout analysis we consider only the leading
order contribution, in $N$, to the free energy density, and in all
coefficients we assume the asymptotic value in $n$ where possible for
brevity. We first write the free energy function making use of the
replica trick~\cite{Mezard:SGT}.
\begin{eqnarray}
 f &=& \lim_{N\rightarrow \infty} -\frac{1}{\beta N} \log
 \sum_{S_i,\ldots,S_N} \exp\{-\beta {\cal H}(S_1,\ldots,S_N)\}\;, \\
&=& \lim_{N\rightarrow \infty} -\frac{1}{\beta N} \lim_{n\rightarrow 0}
 \frac{\partial}{\partial n} \prod_{\alpha=1}^n
 \left[\sum_{S^\alpha_i,\ldots,S^\alpha_N}  \exp\{-\beta {\cal H}
 (S^\alpha_1,\ldots,S^\alpha_N)\}\right] \nonumber \;.
\end{eqnarray}
We are interested in the behaviour of a particular sample drawn
from the ensemble described by the parameters ${\cal I}$. We
anticipate that a sufficient description of any typical instance
will be given by the free energy averaged over instances of the
disorder (self averaging assumption).

After taking the quenched averages of couplings and some
manipulation of the equation form we are able to describe the self
averaged free energy density by
\begin{equation}
\< f \>= \lim_{N\rightarrow \infty} -\frac{1}{\beta N}
\lim_{n\rightarrow 0} \frac{\partial}{\partial n} \int
d{\Phi}d{\hat{\Phi}}
 \exp \{- N \Lambda(n,{\cal I},\Phi,\hat{\Phi}) \}\;, \label{basicsaddlepointform}
\end{equation}
where
${\Phi,\hat{\Phi}}=\{\pi(\vsigma),\pih(\vsigma),\qbaral,\qhbaral,\qalal,\qhalal\}$
are the set of integration variables introduced to allow exact
site factorisation. These may be invoked as order
parameters for various phases.  Considering only leading order
terms in $N$ one can present $\Lambda$ is in the general composite
case of two types of disordered couplings
\begin{eqnarray}
\fl \Lambda &=& - \sum_\alpha \qhbaral \qbaral - \sum_\alal \qalal
\qhalal - \sum_{\vsigma} \pi(\vsigma)\pih(\vsigma) - \log
\Avb{\sum_\vsigma \exp\{\mX\}} \nonumber \\ \fl &-& \frac{\rho}{2}
\sum_{\vsigma_1,\vsigma_2} \pi(\vsigma_1)\pi(\vsigma_2) \SP -
\frac{\beta^2 {\tilde J}}{2} \left[\frac{1}{2} + \sum_\alal
(\qalal)^2\right] \nonumber \\ \fl &-& \frac{\beta \Jo}{2}
\sum_\alpha (\qbaral)^2 - \beta \sum_\alpha \left(h^S \sum_\vsigma
\pi(\vsigma) \sigma^\alpha + h^D \qbaral\right) - \beta^2
\chi^2\left(\frac{1}{2} + \sum_\alal \qalal\right) \;;  \nonumber\\
\fl \mX &=& -\sum_\alpha \qhbaral b \sigma^{\alpha}- \sum_\alal
\qhalal \sigma^{\alpha_1}\sigma^{\alpha_2} - \pih(\vsigma) \;.
\end{eqnarray}
The scaling with $\gamma$ is absorbed in the coupling distribution
$\phi$,$J_0$ and $\Jt$.

From here it is possible to evaluate the integral by the saddle
point method anticipating the large $N$ limit. The self-consistent
equations obtained by the extremisation of the exponent may be
evaluated exactly at high temperature and by a qualified
approximation at lower temperature. The saddle point equations for
the order parameters are
\begin{eqnarray}
\fl  \qbaral &=& {\cal N} \Avb{\sum_\vsigma b \sigma^\alpha \exp{\mX}}
\;;\qquad \qhbaral = -\beta \Jo \qal - \beta h^D \label{qbarsaddle}\\
\fl  \qalal &=& {\cal N} \Avb{\sum_\vsigma \sigma^{\alpha_1} \sigma^{\alpha_2} \exp{\mX}}
\;;\qquad  \qhalal = - \beta^2 \Jt \qalal - \beta^2 \chi^2 \label{q2saddle}\\
\fl \pi(\vsigma) &=& {\cal N} \Avb{\exp{\mX}} \nonumber
\;;\\ \fl  \pih(\vsigma_1) &=& - \rho \sum_{\vsigma_2}
\pi(\vsigma_2)\SP
 - \beta h^S \sum_{\alpha_1} \sigma^{\alpha_1} \label{pisaddle} \\
\fl {\cal N} &=& \Avb{\sum_{\vsigma} \exp{\mX}}^{-1} \nonumber
\end{eqnarray}
from which the conjugate order parameters $\hat{\Phi}$ may be
eliminated. This representation is convenient for making
general ansatze~\cite{Monasson:OP} on the order parameters in
(\ref{RSassumption}).

\subsection{Alternative formulation}
It is useful to represent the exponent (\ref{Lambda1}) by choosing a
standard parameterisation for $\pi$ allowing components to be
associated with different types of order. One can equivalently
generate this representation directly by choosing different order
parameters and expansions in the replica calculation, the
representation remains a standard one~\cite{Viana:PD,wong_b}.  Working
at the saddlepoint we are able to substitute saddlepoint definitions
for $\pih$,$\qhbaral$ and $\qhalal$ to obtain an equation in only
$\pi(\vsigma)$,$\qbaral$ and $\qalal$.  We then reparameterise $\pi$,
 the generating function of the order parameters, by the complete
expansion
\begin{equation}
\pi(\vsigma)=1 + \sum_\alpha \qal \sigma^\alpha + \sum_\alal \qalal
\sigma^{\alpha_1}\sigma^{\alpha_2} + \sum_{l=3}^n q_{\< \alpha_1
\ldots \alpha_l\>} \sigma^{\alpha_1} \ldots
\sigma^{\alpha_l}\;,\label{piexp}
\end{equation}
The components $\qs$ as well as $\qbaral$ represent spin-spin
correlation order parameters between replica~\cite{Nishimori:SP}. In
the case $\bbar=1$ the distinction between $\qal$ and $\qbaral$ is
artificial, as emerges from the calculation. We can rewrite the
exponent at the saddlepoint as
 \bea \fl\Lambda &=& - \log \Avb{\sum_\vsigma \exp\{\mX\}} +
\frac{\beta J_0}{2} \sum_\alpha (\qbaral)^2 + \frac{\beta^2 \Jt +\rho
\mT_2}{2} \sum_\alal (\qalal)^2 \nonumber\\ \fl &+& \frac{\rho
\mT_1}{2} \sum_\alpha (\qal)^2 + \sum_{l=3} \frac{\rho \mT_l}{2}
\sum_{\<\alpha_1 \ldots \alpha_l\>} (q_{\alpha_1 \ldots \alpha_l})^2
\;.\nonumber\\ 
\fl \mX &=& \sum_\alpha \beta ( b (\Jo \qbaral + h^D) + \rho
\mT_1 \qal + h^S) \sigma^{\alpha} + \beta^2\sum_\alal ((\Jt + \rho
\mT_2)\qalal + \chi^2) \sigma^{\alpha_1}\sigma^{\alpha_2} \nonumber \\
\fl &+& \rho \sum_{l=3} \mT_l \sum_{\<\alpha_1 \ldots \alpha_l\>}
q_{\alpha_1 \ldots \alpha_l} \sigma^{\alpha_1} \ldots
\sigma^{\alpha_l}\;. \label{Lambda2} \eea
up to addition of constants. The quantities $\mT_i$ are to leading
order in $n$
\begin{equation}
 \mT_i= \int d\phi(x) \tanh^i(\beta x) \,
\end{equation}
with $\phi$ being the coupling distribution. The new order
parameters must satisfy the set of coupled saddlepoint equations
\begin{eqnarray}
 q_{\langle\alpha_1 \ldots \alpha_p\rangle} &=& \Avb{ {\cal N}\sum_\vsigma
 \{\sigma^{\alpha_1} \ldots \sigma^{\alpha_p}\} \exp{\mX}}
 \label{qsaddle}\;,
\end{eqnarray}
with (\ref{qbarsaddle}) still applying.  Non-zero order solutions of many types may exist but
solutions are non-trivial except at high temperature and zero exteral fields. In the next section we examine solutions for vanishing
external fields $h_S=h_D=\chi=0$ emergent in the high temperature
regime.

\section{High temperature solutions at zero external field}
\label{sec:highT}

In the case of zero external fields there exist phases classified
by the set of non-zero order parameters. For the low $\beta$
regime the unique stable solution to these equations is the
paramagnetic one (P), with all order parameters zero (\ref{qsaddle}). It is possible to determine a hierarchy of
critical temperatures~\cite{Viana:PD} for the emergence of
different types of non-paramagnetic order solution by writing the
saddle point equations (\ref{qsaddle}) in
cases with all but one type of order parameter types zero. A single
spin (ferromagnetic, F) order emerges subject to a non-zero
solution of
\begin{equation}
\left(\begin{array}{c}
\qbaral \\ \qal \end{array}\right) = \left(\begin{array}{c}\tanh(\beta
J_0 \qbaral) + \bbar \tanh(\rho \mT_1\qal) \\ \bbar \tanh(\beta J_0
\qbaral) + \tanh(\rho \mT_1 \qal)\end{array}\right) \label{sce1}\;.
\end{equation}
A Spin-Glass (SG) order requires a non-zero solution to
\begin{equation}
 \qalal = \tanh((\beta^2 \Jt + \rho \mT_2)\qalal) \label{sce2}\;.
\end{equation}
Due to the convexity of the tanh function the criteria
is described equivalently by a linear expansion
\begin{equation}
 \left[\lambda^S = \beta^2 \Jt + \rho \mT_2\right] > 1\label{2spinorder}\;.
\end{equation}
Higher order solutions, indicated by $q_{\langle\alpha_1 \ldots
\alpha_p\rangle}$ emerge subject to the criteria $\rho\mT_p>1$, by
a similar linearisation to (\ref{2spinorder}). However, no
solution of this type can emerge at higher temperature than that
indicated by equation~(\ref{2spinorder}) as the following
inequality holds for arbitrary coupling distribution
$\phi$~\cite{Viana:PD}
\begin{equation}
 (\beta^2 \Jt + \rho \mT_2) \geq \lambda^{(p)}\;, \label{pspinorder}
\end{equation}
defining $\lambda^{(p)}=\rho\mT_p$ for $p>2$, the inequality becomes strict in the case $\Jt>0$.

One can apply similar arguments in convexity to aid solution
finding in the second case (\ref{sce1}), we will restrict
attention to the cases $\bbar=1$ and $\bbar=0$. In the first case
the right-hand-side of (\ref{sce1}) is identical in the two
components so that the order emerges uniquely along the direction
$\tal\propto\qal=\qbaral$. The
criteria for a non-zero solution is
\begin{equation}
 \left[ \lambda^F_+ = \beta J_0 + \rho \mT_1 \right] > 1\label{1spinorder1}\;.
\end{equation}
For the case $\bbar=0$ the situation is also relatively simple, by the
convexity properties of the $\tanh$ it is sufficient to linearise
(\ref{sce2}) in $\qbaral$, $\qal$. The independent processes can then
yield solutions depending on one of two eigenvalues meeting the criteria
\begin{equation}
    \left[ \lambda^F_\pm = \frac{1}{2}\left( \beta J_0 + \rho \mT_1 \pm
    \sqrt{(\beta J_0 - \rho \mT_1)^2 + 4 \bbar \beta J_0 \rho
    \mT_1}\right)\right] > 1\;, \label{1spinorder}
\end{equation}
written to be inclusive of both cases $\bbar=0$ and $\bbar=1$
(\ref{1spinorder1}). Except in the case of a critical temperature with $\rho \mT_1=\beta J_0=1$
the largest eigenvalue $\lambda^F=\lambda^F_+$ will determine the
type of the emergent one spin order. We use the generalised order parameter $\tal\propto v_1 \qbaral + v_2 \qal$ to correspond to the mode $\lambda^F_+$, $\vv$ depending on the
type of Ferromagnetic order. We assume throughout the paper that
the discrete symmetry in solutions $\pm\tal$ is broken by a small
external field ($h_S$ or $h_D$) aligned with the positive
solution. When $\lambda^F_+ \approx \lambda^F_-$ one must consider both
$\qal$ and $\qbaral$ becoming non-zero simultaneously, the
consequences of generalising the following sections' analysis to
include this case are considered in ~\ref{app.SAUS}.

Equations (\ref{2spinorder}),(\ref{pspinorder}) and
(\ref{1spinorder}) indicate that for any mixed system ($\Jt$ or
$\Jo \neq 0$) there is a transition at some temperature towards a
ferromagnetic or SG phase. The effect of $\rho$, which indicates a
percolation in the sparse couplings only if $\rho>1$, generates no
obvious criticality in the expressions, except in cases where
$\Jo\leq 0$ ($\Jt=0$) then $\rho$ must exceed $1$ for the
possibility of a ferromagnetic (SG) high temperature transition,
regardless of bond strength or distribution in the sparse part. In
this scenario of negative mean coupling in the dense subsystem, $\lambda^F_+$ (\ref{1spinorder}) is a convex, not monotonically
increasing, function of $\beta$. This means for example that by
continuous variation of several ensemble parameters one is able to
create a first order Paramagnetic-Ferromagnetic (sparsely aligned)
transition, with the upper critical temperature changing
discontinuously in $\gamma$. Details may depend on the
sparse coupling distribution $\phi$.

There exist a number of alternative rigorous methods to attain
such a set of paramagnetic high temperature transition points
without the use of replicas. In particular we note the
results~\cite{Ostilli:ISG}, which allow the high temperature
transitions to be found by transformation of a disordered Ising
spin model to a uniform interaction Ising model for many
topologies, thereby allows easy identification of transition
points.

\subsection{High temperature auxiliary system}
\label{sec.HT} We generate a system to describe the
non-Paramagnetic phase in the vicinity of the high temperature
transition with either or both $\Delta_1=\lambda^F_+-1$ or
$\Delta_2=\lambda^{SG}-1$ small and positive, and with $\lambda^{(p)} < 1$ and $\lambda^F_- < 1$ (there is only one
potentially non-zero ferromagnetic alignment). In this scenario we
may describe the leading order behaviour to second order in
$\Delta$ exactly by an auxiliary saddlepoint equation - which may
be subject to exact analysis. This approach is motivated by
related methods for sparse SG~\cite{Viana:PD,Mottishaw_stability}.

We first introduce some important definitions chosen to be
compatible with~\cite{Viana:PD}
\begin{eqnarray}
\fl r_1 &= \frac{1}{\lambda^F_+} - 1 \label{r3r4}\;;
 \qquad r_2=\frac{1}{\lambda^S} - 1\;;
 \qquad  r_{i>2}&=\frac{1}{\lambda^{(i)}} - 1  \;; \nonumber \\
\fl \tal &= \lambda^F_+ (v_1 \qal + v_2 \qbaral) + \beta (v_1 h^S +v_2 h^D) ;\qquad \; & Q_{(\alpha_1 \alpha_2)} =
 \lambda^S \qalal + \beta^2 \chi^2 \label{fsgop}\;; \\
 \fl Q_{(\alpha_1 \alpha_2 \alpha_3)} &= \lambda^{(3)} q_{\alalal}\;;\qquad & Q_{(\alpha_1 \alpha_2 \alpha_3 \alpha_4)} = \lambda^{(4)} q_{\alalalal} \nonumber
\end{eqnarray}
Notation $()$ is to indicate any permutation in the set of indices, the new order parameters are
identical for any ordering of indices. The notation $()$ is dropped for abbreviation in the parameters henceforth.
The components of $\vv=(v_1,v_2)$ describe the different possible
alignments of the ferromagnetic order at the high temperature
boundary: $(1,1)$ for $\bbar=1$ and $(1,0)$ $[(0,1)]$ for
$\bbar=0$ with $\rho \mT_1 < [>] \beta J_0$, respectively.

At a certain temperature, the minimum $\beta$ which satisfies
either (\ref{2spinorder}) or (\ref{1spinorder}), a new phase
continuously emerge from the paramagnetic solution which is either
ferromagnetic or SG, respectively. A sufficient description of these solutions
in the vicinity of this high temperature boundary is given by an
expansion in the restricted set of order parameters: ferromagnetic
$\tal$ and SG $Q_{\alpha_1 \alpha_2}$ under some ansatz. The rescaling of order
parameters (\ref{fsgop}) is to abbreviate the
expressions, and we consider the case of negligable external fields in
examining the transitions. For $\lambda^X<1$ the saddle point solution for the $X^{th}$ order parameter is zero unless there is a non-zero component in the higher order parameters, we say the order is induced. For example a non-zero $\tal$ induces order in $\Qalal,\Qalalal$ when $\lambda^S,\lambda^{(3)}<1$, and $\Qalal$ induces order in $\Qalalalal$ when $\lambda^{(4)}<1$. We include order parameters upto $4^{th}$ order, this set is sufficient to describe the leading order behaviours of the phases. These inclusions discriminate the approach from a simple mean field (fully connected) approximation to the correlation structure of the sparse subsystem.

Calculations of the significant higher order parameters through
the saddle point equations (\ref{qsaddle}) and of $\Tr \<\exp
\mX\>$ are undertaken in~\ref{app.VB}. The auxiliary expression for
$\Lambda$~(\ref{Lambda1}) can then be found by expansion in
$\tal$,$\Qalal$,$\Qalalal$,$\Qalalalal$ to significant order as
\begin{eqnarray}
\fl \Lambda &=& \frac{1}{2}\sum_\alpha \left( (t_\alpha -\beta h)^2 /\lambda^F_+
\!-\! (t_\alpha)^2 \right) \!+\! \frac{1}{4}\sum \left( (Q_{\alpha_1 \alpha_2} -\beta^2 \chi^2)^2 /\lambda^S
- (Q_{\alpha_1 \alpha_2})^2 \right) \!+\! \frac{r_3}{12}\sum \Qalalal^2\nonumber\\
\fl &+&  \frac{r_4}{48} \Qalalalal^2  +\frac{1}{12}\sum \tal^4 + \frac{1}{2!12 }\sum Q_{\alpha_1 \alpha_2}^4 - \frac{1}{45}\sum \tal^6 \label{Lambda3} \\
\fl &-& \frac{1}{2!}\sum t_{\alpha_1}t_{\alpha_2} Q_{\alpha_1 \alpha_2} + \frac{2}{2! 3 }\sum t_{\alpha_1}^3 t_{\alpha_2} Q_{\alpha_1 \alpha_2} + \frac{1}{2! 3 }\sum t_{\alpha_1} t_{\alpha_2} Q_{\alpha_1 \alpha_2}^3 - \frac{3}{3!}\sum t_{\alpha_1} t_{\alpha_2} Q_{\alpha_2 \alpha_3} Q_{\alpha_1 \alpha_3} \nonumber\\
\fl &-& \frac{1}{3!} \sum Q_{\alpha_1 \alpha_2} Q_{\alpha_2 \alpha_3} Q_{\alpha_1 \alpha_3} \!-\! \frac{12}{4!}\sum t_{\alpha_1} t_{\alpha_2}\! Q_{\alpha_1 \alpha_3} \!Q_{\alpha_2 \alpha_4}\! Q_{\alpha_3 \alpha_4} \!-\! \frac{3}{4!} \sum Q_{\alpha_1 \alpha_2} \!Q_{\alpha_1 \alpha_3} \! Q_{\alpha_2 \alpha_4} \! Q_{\alpha_3 \alpha_4}\nonumber\\
\fl &-&  \frac{3 v_2}{3!}\sum t_{\alpha_1} Q_{\alpha_2 \alpha_3}\Qalalal - \frac{v_2}{3!}\sum t_{\alpha_1} t_{\alpha_2} t_{\alpha_3} \Qalalal - \frac{3}{4!}\sum \Qalalalal Q_{\alpha_1 \alpha_2} Q_{\alpha_3 \alpha_4} \nonumber \;, 
\end{eqnarray}
where summations are over the {\it unordered distinct} indices.
The difference in this expression from a simple combination of two fully or sparsely connected systems is in the temperature dependence of terms $\Delta_1$, $\Delta_2$ and in the entropic terms of coefficient $v_2$ -- the sparsely induced ferromagnetic/mixed solutions differs at second order.

The simplest solution to this equation, consistent with a non-zero
solution and the  $n\rightarrow 0$ limit, is a replica symmetric
one plus fluctuations: taking $\tal=t+S_\alpha$, $\Qalal=Q+R_{\alpha_1 \alpha_2}$, $\Qalalal=Q_3+\delta\Qalalal$ and $\Qalalalal=Q_4+\delta \Qalalalal$. These fluctuations should be considered independent upto permutations on the set of indices.
We may now rewrite the auxiliary free energy (\ref{Lambda3}) in terms of the RS solution and fluctuations as
\begin{eqnarray}
  \fl\Lambda &=& \Lambda(t,Q,Q_3,Q_4) + \mW \sum S_\alpha + \mZ \sum R_{\alpha_1 \alpha_2} + \mZ_3 \sum \delta \Qalalal + \mZ_4 \sum \delta \Qalalalal
  \label{LambdaAT} \\
\fl &+& \mA \sum (S_\alpha)^2 + \mB \sum S_\alpha S_\beta  \!+\! 2 \mC \sum R_{\alpha \beta} S_\alpha + \mD \sum R_{\alpha \beta}
S_\gamma \!+\! \frac{\mP}{2}\sum R_{\alpha \beta}^2
\!+\! \mQ \sum R_{\alpha \beta} R_{\alpha \gamma} \nonumber\\
   \fl &+&  \frac{\mR}{4}\sum R_{\alpha \beta} R_{\gamma \delta} \!+\! \sum \delta Q_{\alpha \beta \gamma} (X \delta Q_{\alpha \beta \gamma} \!+\! Y  R_{\alpha \beta} \!+\! Z S_\alpha) \!+\! \sum \delta Q_{\alpha \beta \gamma \delta} (X_4 \delta Q_{\alpha \beta \gamma \delta} \!+\! Y_4 R_{\alpha \beta}) \nonumber\;.
\end{eqnarray}
The prefactors are chosen to make connection with a standard stability analysis result which we use in the following section~\cite{Almeida:SSK}. For the RS solution to be a saddlepoint solution the terms linear in the fluctuations must vanish so that:
\begin{eqnarray}
\fl \mW &=0 &= \frac{-\beta h}{\lambda^F_+} + t\left(r_1  + Q  + \frac{1}{3} t^2 -
 2 Q^2  - \frac{2 }{15} t^4 - \frac{4}{3}Q t^2 + \frac{17}{3} Q^3 - v_2 \frac{Q_3}{t}(t^2 + Q)\right)\; \\
\fl \mZ &=0 &= \frac{-\beta^2 \chi^2}{2 \lambda^S} + \frac{1}{2} r_2 Q + Q^2 - \frac{1}{2} t^2 + 2 Q t^2 - \frac{17}{6} Q^3 + \frac{1}{3}t^4 - \frac{17}{2} Q^2 t^2 + v_2 Q_3 t - \frac{3}{2} Q_4 Q, \\
\fl \mZ_3 &=0 &= \frac{v_2}{r_3}(v_2 t^3 + 3  Q) \qquad, \qquad
\mZ_4 =0 = \frac{3 Q}{r_4}
\label{higherorderparams}
\end{eqnarray}
These allow paramagnetic ($t=Q=0$), ferromagnetic ($t\neq
0$,$Q\neq0$) and SG ($t=0$,$Q\neq0$) solutions.

\subsection{Stability near the high temperature transition}
\label{sec.VB}

The RS solution has $t$ and $Q$ such that the exponent is a local
extremum. The saddlepoint is a stable local maxima if the
quadratic form, which may be described by a real symmetric
Hessian, is positive definite. This is difficult to calculate in
the general case with non-zero fluctuations in all the order
parameters. Instead we wish to consider a restricted set of
fluctuations with $\delta \Qalalal=\delta \Qalalalal = 0$, thus we
are testing coupled instabilities with respect to marginal
magnetisations and spin-spin correlations only. This restricts
somewhat the set of possible perturbations but provides a concise
approximation to the stability properties, including the Replica
Symmetry Breaking (RSB) ansatz, we expect to generalise very well
to inclusion of more complicated multispin instabilities.

We
can in the restricted case calculate the eigenvalues of the quadratic form which are
degenerate and only of three types in the limit $n \rightarrow
0$~\cite{Almeida:SSK}. The stabilities depend on a combination of
coefficients given by
\begin{eqnarray}
 \mA' &=& \mA - \mB = \frac{r_1}{2} + \frac{1}{2}Q + \frac{1}{2}t^2 - Q^2 - \frac{1}{3}t^4 + \frac{17}{6}Q^3 + \frac{v_2}{r_3}\left(\frac{3}{2} Q t^2+\frac{v_2}{2} t^4\right)\label{termS}\\
 \mC' &=& \mC - \mD  = t\left( -\frac{1}{2} -\frac{17}{2} Q^2 + 2 Q +\frac{2}{3}Qt^2 + \frac{v_2}{2 r_3}\left( 3 Q + v_2 t^2\right)\right)
 \label{termSQ}\\
 \mP &=& \frac{r_2}{2} + t^2 + \frac{1}{2}Q^2 \label{termQQ}\\
 \mQ &=& -\frac{Q}{2} - \frac{t^2}{2} + \frac{3}{2}Q^2 + 3 Q t^2\label{termQQ2}\\
 \mR &=& - \left(\frac{3}{2 r_4} + 1\right) Q^2 - 2 t^2 Q,
\end{eqnarray}
where we have substituted the definitions (\ref{higherorderparams}) for the higher order parameters. These combine to give the longitudinal instabilities
\begin{equation}
 \lambda_\pm = \frac{1}{2}\left(\mA'+\mP-4\mQ+3 \mR\pm \sqrt{(\mA'-\mP + 4\mQ - 3 \mR )^2 -
  8 (\mC')^2}\right)\;, 
\end{equation}
and the replicon instability
\begin{equation}
 \lambda_1 = \mP - 2 \mQ + \mR \;.
\end{equation}

\begin{figure}
\begin{center}

\includegraphics*[width=\linewidth]{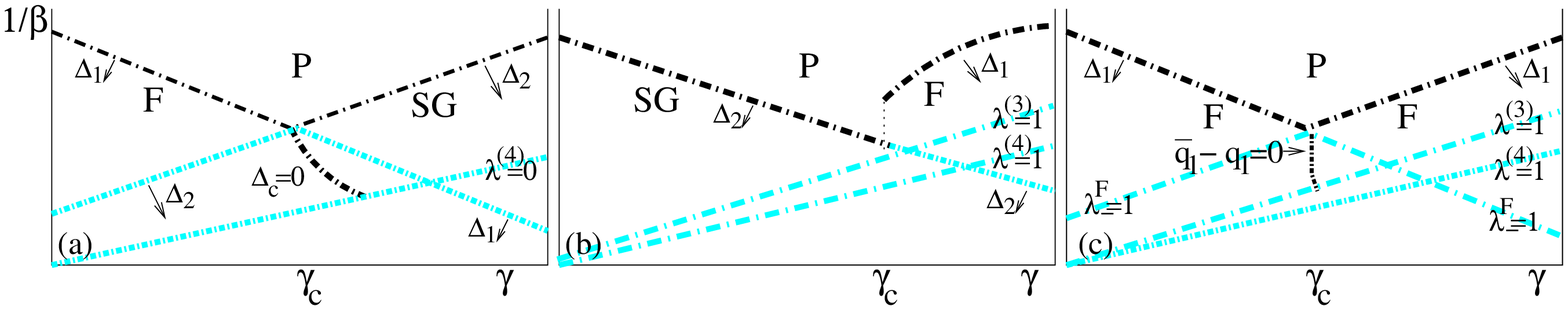}
\caption{A qualitative description of RS solutions and their
instabilities, the auxiliary system results being valid in the vicinity of the 
high temperature transition line.  Dot-dashed lines indicate emergence of high temperature 
phase transition. The darker (upper) lines indicate the relevant high temperature solution from this set.
Left - The high temperature
transition can be of ferromagnetic or SG type. If the types of
order in the sparse and dense part are not complementary there can
exist a triple-point ($\gamma_C$). Below the triple-point the type
of order will in most cases be that induced by the dense
subsystem. The additional dot-dashed line descending from the triple-point corresponds to
the RS F-SG transition line (\ref{criticalline}), the RS
ferromagnet being the stable solution between this line and the
corresponding Ferromagnetic high temperature transition. The triplepoint analysis is valid where $\lambda^{(p)}<1$. 
One can reverse the
labelling F,SG and corresponding $\Delta$ symbols in the diagram,
with the qualitative nature of the diagram being similar. Centre
- In cases where the dense distribution has \emph{negative} mean
coupling the ferromagnetic solution to (\ref{1spinorder}) may
disappear discontinuously, so that the high temperature transition
is a discontinuous one, the triple-point analysis applies to no
part of the diagram. Right - In the case of competing alignments
of ferromagnetic order then there will exist competition between
these alignments; again at a triple-point, the solution below this
point can have $(\qal,\qbaral)$ non-zero in a combination
determined by~(\ref{sce1}), the free-energy is symmetric about a dotted line $\beta J_0 = \rho \mT_1$ at leading order in $\Delta$. This is a vertical line in the $\beta-\gamma$ plane at leading order. \label{fig:ER}}
\end{center}
\end{figure}

By an expansion of the coefficients, in terms of
$\Delta_2 = \lambda^S - 1$ near the SG boundary and $\Delta_1 =
\lambda^F_+ - 1$ in the vicinity of a ferromagnetic boundary, it is
possible to determine leading order RS solutions and their
stability properties. We consider the two cases that one component
is small and the other large, and the case of a fine balance
between the two ($\Delta_1= O(\Delta_2)$). Results should be
interpreted with reference to figure~\ref{fig:ER}.

\subsection{Solutions below the high temperature transitions}
\subsubsection{Paramagnetic (P) solution.}
The P solution ($t=0,Q=0$) is unstable everywhere below
the high temperature transition point and stable everywhere above
it; eigenvalues being proportional to $-\Delta_1$ and $-\Delta_2$.

\subsubsection{Ferromagnetic (F) solution.}
In the regime where $\Delta_2 \ll 0$ and $\Delta_1 > 0$,  only the
F and P solutions exist, one finds
the F solution to leading order has $r_2 Q = t^2 =
3 r_2 \Delta_1/(3 + r_2)$. 
The longitudinal ($\lambda_+$), and replicon
eigenvalues are proportional to $r_2$, which is positive in
this regime. The other longitudinal eigenvalue is proportional to
$\Delta_1$, which is positive and coincident with the high
temperature boundary.

\subsubsection{Spin Glass (SG) solution.}
In the regime where $\Delta_1 \ll 0$ and $\Delta_2 > 0$ only the SG and P solutions exist. The paramagnetic
solution is unstable in a longitudinal mode corresponding to the
high temperature boundary. One finds the SG solution gives
$Q=\Delta_2/2,t=0$ to leading order. The longitudinal eigenvalue
$\lambda_+$ is coincident with $r_1$, which is positive. The
other longitudinal eigenvalue is proportional to $\Delta_2$, which
is positive. The replicon instability is given to leading order as
\begin{equation}
 \lambda_1 = \Delta_2^2 \left(-\frac{3}{8} - \frac{1}{6 r_4} + \frac{9}{8 r_4^2}\right), \label{SGrepmode}
\end{equation}
which may be negative or positive depending on the value $r_4$. The expected replica symmetry breaking instability is attained when $\Jt$ or $\rho$ is large, but where $\rho$ is near the percolation threshold and with $\Jt$ small the spin glass can be stable. Since our analysis includes the VB model as a special case, which is prooven to be unstable in the RS spin glass solution ~\cite{Mottishaw_stability}, this result must indicate some pathology in the stability analysis -- the failure to consider higher order fluctuations. Nevertheless our results may be indicative of a general weakening of instabilities as the sparse coupling limit is approached.

\subsubsection{Co-emergence regime $\lambda^S\approx\lambda^F$.}
In this case we make an expansion with both
$\Delta_1$ and $\Delta_2$ small, and consider both the SG and
F solution. The SG solution and eigenvalues are
unchanged at leading order except in
\begin{equation}
\fl \lambda_- = \frac{\Delta_1}{8}\left(3 \mu - 2 - \sqrt{4 + 4 \mu +
\mu^2}\right)+\Delta_1^2\left(\frac{9}{4 r_4^2} + \frac{5}{12 r_4} - \frac{1}{2} + O(2 -\mu)\right) \label{SGlongmode}
\end{equation}
where $\Delta_2 = \mu \Delta_1$. This coefficient is negative provided $0<\mu<2$, so a longitudinal instability emerges when
\begin{equation}
 \left[\Delta_c = \Delta_1 - \Delta_2/2 \right]= 0 \;. \label{criticalline}
\end{equation}

The ferromagnetic solution is changed in both the RS order parameters
\begin{equation}
  t^2 = 2\Delta_1^2 - \Delta_1 \Delta_2 \qquad ; \qquad Q=\Delta_1 + \Delta_1^2\left(\frac{1+\mu}{3}+\frac{3 v_2}{r_3}\right)
\end{equation}
at leading orders, and in eigenvalues
\begin{eqnarray}
\lambda_- &=&  \Delta_1 \frac{2-\mu}{2} + \Delta_1^2 \left(\frac{3 v_2}{2 r_3}\right)+ \Delta_1^2 O(2-\mu) \label{lambdaminuscomplicated}\\
\lambda_1 &=& \Delta_1 \frac{2-\mu}{2} + \Delta_1^2
\left(-\frac{1}{2} - \frac{3}{2 r_4} +\frac{3 v_2}{r_3} +
\frac{4(\mu-2)}{3} + \frac{(\mu-2)^2}{2}\right)
\label{lambda1_complicated}
\end{eqnarray}
The other eigenvalue remains coincidence with $\Delta_1$ for
all $\mu$ (indicating stability). Thus at leading order we have a
transition from an $RS$ ferromagnetic system to a
longitudinally stable SG when $\mu=2$ (\ref{criticalline}).

Consider first the situation in which at the triple point $r_3$ is not sufficiently small in comparison with $r_4$ to make (\ref{lambda1_complicated}) positive, or that the alignment of the ferromagnetic moment is in the dense part $v_2=0$. The second order corrections to the eigenvalues for the ferromagnetic regime indicate that in the vicinity of the transition the replicon mode (\ref{lambda1_complicated}) is more negative than the non-negative valued longitudinal mode (\ref{lambdaminuscomplicated}). We have a negative value for the replicon instability, but positive value for the longitudinal stability on the critical line, so that at second order we can predict the existence of a RSB unstable phase of non-zero magnetic moment (a mixed phase, M) between the ferromagnetic and SG solutions in a region about the critical line. This result is qualitatively similar to the Viana Bray (sparse) model stability result for variation of $\rho$~\cite{Viana:PD}. If the replicon eigenvalue is positive on the critical line the mixed phase may be shifted marginally about the critical line. Unfortunately the term $r_3$ is sufficiently small to cause positivity in the second order term for many ensembles. A similar complicated dependence of the longitudinal mode exists for the spin glass solution \ref{SGlongmode} -- the results in combination suggest existence of some systems with longitudinally stable F-SG coexistence regimes, rather than a mixed state, near the critical line. However, we suspect these details to be artefacts of the restricted stability analysis.

\subsection{Concluding remarks on high temperature solutions}
The RS ferromagnetic solution is the unique stable RS solution
where it exists, except near the SG transition point (\ref{criticalline}).
The SG RS solution is we suspect an unstable one in the replicon mode, although the 
stability analysis indicates that at higher temperature there may be a region in which it 
is an RS stable solution.
The paramagnetic solution
is unstable everywhere below the high temperature transition
lines. The ferromagnetic phase becomes unstable to replica symmetry breaking
(in $\{\tal,\Qalal\}$) in the vicinity of the transition, generating a mixed phase.

It is of course essential to consider the line
(\ref{criticalline}) represented in terms of some variation of our
parameters. This line may be calculated to leading order as a
function of $\delta \gamma$ and $\delta \beta$, which is
undertaken in \ref{app.TP}. In so doing we find the line is
typically orientated towards $\gamma$ greater than $\gamma_C$ as
temperature is lowered, indicating a transition from 'dense' to
'sparse'-type order; a result which is applicable for either type
of order (SG or F) in the dense part. This result also implies,
interestingly, that ergodicity breaking may disappear as
temperature is lowered; a sufficient dense-coupling tendency
towards ferromagnetism may allow such a transition in the case of
weakly ordered sparse SG. One can consider higher order
corrections to the eigenvalues in the vicinity of the
triple-point, and in so doing we expect that in most cases a mixed
solution will be found, that in some region of parameter space the
RS ferromagnetic solution will be unstable when second order
corrections are included.

This analysis leaves open the corrections to the stability
analysis due to higher order parameters, and results at second order:
(\ref{SGlongmode}),(\ref{lambda1_complicated}) and (\ref{SGlongmode}), motivate such a consideration. A
sensible way to probe higher order parameters might be to consider
other restricted sets of fluctuations, $\delta \Qalalal$ a function
of $\delta \tal$ and $\delta \Qalal$ for example. We have found that complicated terms in
$r_3,r_4$ appear also in these cases, but we anticipate increased instability if
fluctuations are allowed in all 4 order parameter types. It is likely
such an analytic effort would however be better directed to a more
general 1RSB formulation solved by numerical methods. The case of
a degenerate transition ($\lambda^F_- \approx \lambda^F_+$) by a
comparable method is considered in
\ref{app.SAUS}.

\section{Low temperature results by RS assumption}
\label{sec:lowT}

\subsection{Replica symmetric ansatz}
\label{RSA}
The RS ansatz is here applied to the full free energy expression, the order
parameters being invariant with respect to relabelling of spins.
Returning to our order parameter description in terms of
$\pi(\vsigma)$ we have the standard definitions
\begin{equation}
\fl \pi(\vsigma)
= \int dW(h) \frac{\exp\{\beta h \sum_\alpha \sigma_\alpha
\}}{2\cosh(\beta h)} \qquad;\qquad
\qbaral = \qbaro \qquad;\qquad
\qalal = q_2  \label{RSassumption}
\end{equation}
so that $W(h)$ is now a normalised
distribution on the real line describing the moments of
$\pi(\vsigma)$, which must be determined. The self consistent
equations may then be reexpressed in terms of the set of
parameters $W,\qbaro,q_2$ at the saddlepoint
\bea
\fl W(h) &\propto& \< \prod_{i=1}^L \left[\int dW(h_i) d\phi(x_i)\right]
\delta(h - H) \>_{b,L,\lambda_1,\lambda_2}\label{Wrs} \\
\fl \qbaro &\propto& \int d h \tanh (\beta h) \< b \prod_{i=1}^L \left[\int
dW(h_i)
d\phi(x_i)\right] \delta(h - H)\>_{b,L,\lambda_1,\lambda_2} \label{qbar1rs}\\
\fl q_2 &\propto& \int d h \tanh^2 (\beta h) \< b \prod_{i=1}^L \left[\int
dW(h_i)
d\phi(x_i)\right] \delta(h - H)\>_{b,L,\lambda_1,\lambda_2} \label{q2rs}\\
\fl H &=&  b \Jo \qbaro + b h^D + h^S + \lambda_1 \sqrt{\Jt q_2} + \lambda_2
\chi +
 \frac{1}{\beta}\sum_{i=1}^L \atanh(\tanh(\beta x_i) \tanh(\beta h_i))
\eea
The average in $b$ is with respect to $\bbar$, $\lambda$ averages
are over Gaussian distributions of zero mean and unit variance,
and $L$ average is over a Poissonian probability distribution of
parameter $\rho$. For $\bbar=0$ one may reduce the expression
(\ref{qbar1rs}) to a simple one, as a moment of the distribution
$W(h)$.

This equation may be solved by the standard method of population
dynamics~\cite{Mezard_popdynamics}, the distribution $W(h)$ is
represented at time $t$ by a histogram of $N$ fields $\{h_i\}$,
alongside the scalar order parameters $\qbaro^{(t)}$ and
$q_2^{(t)}$. During any update step a single field ($i$) is
randomly selected and updated according to a sample from the
quenched parameters $\{L,b,\vx,\vh,\lambda_1,\lambda_2\}$:
\bea
  h_i^{(t+1)} &=& b^{(t+1)} \Jo \qbaro + b^{(t+1)} h^D + h^S + \lambda_1^{(t)}
 \sqrt{\Jt q_2} + \lambda_2^{(t)} \chi \nonumber \\&+&
 \frac{1}{\beta}\sum_{j=1}^L \atanh(\tanh(\beta x^{t}_{i_j}) \tanh(\beta h^{t}_{i_j}))\label{fieldupdateline1}\\
  h_{j \neq i}^{(t+1)} &=& h_{j \neq i}^{(t)} \qquad\;:\qquad b_{j \neq
i}^{(t+1)}
  = b_{j \neq i}^{(t)} \;,
\eea
all other fields being left invariant. Following this we must
update the other order parameters, the field $h_i^{(t)}$ is
removed and the new field $h_i^{(t+1)}$ is added to obtain
\bea
 \qbaro^{(t+1)} &=&  \qbaro^{(t)} + b^{(t+1)} \tanh(\beta h_i^{(t+1)}) -
  b^{(t)} \tanh(\beta h_i^{(t)}) \nonumber \\
q_2^{(t+1)} &=& q_2^{(t+1)} + \tanh^2(\beta h_i^{(t+1)}) -
\tanh^2(\beta h_i^{(t)}) \ . \label{eq:saddleq}
 \eea
Note that in the case that $\bbar\neq 1$ it is necessary to
associate with each field in the histogram the sampled disorder
$b^{(t)}$ in the relevant field update step, in order that
$\qbaro^{(t+1)}$ may be incrementally updated.

Convergence through this procedure to the correct solution, to
within numerical accuracy dependent on $N$, is fairly robust. In
order to avoid systematic errors, and attain convergence in a
suitable time scale we must carefully choose initial conditions
and population size, decide upon convergence criteria, and a
sufficient level of sampling, as discussed in~\ref{app.PD}.

The non-variational free energy $\beta \<f\>$, coincident with
$\Lambda$ after the appropriate limits have been taken, may be
written as
\begin{eqnarray}
\fl  \frac{\partial}{\partial n} \Lambda &\doteq& -\frac{\rho}{2}\<\log \cosh
\beta x\>
+ \frac{\rho}{2}\<\log(1 + \tanh\beta x \tanh\beta h_1 \tanh \beta h_2)\>
- \<\log 2 \cosh (\beta h)\> \nonumber \\
\fl &+& \rho\<\log \cosh \atanh(\tanh(\beta h)\tanh(\beta x))\> +
\frac{1}{2} \beta J_0 \qbaro^2 -\frac{1}{4}\beta^2 \Jt (1-q_2)^2 -
 \frac{1}{2}\beta^2 \chi^2 \;.\label{f}
\end{eqnarray}
With appropriate scaling of the coupling distribution $\phi$ and
$\gamma$ with $N$ it is possible to show equivalence of the sparse
part with the dense part at large connectivity $\rho$.

Other quantities of interest are
\begin{equation}
\< S^{\alpha_1} \ldots S^{\alpha_k}\> = \int dW(h) \tanh(\beta h)\;,
\label{locfielddist}
\end{equation}
where $W(h)$, the auxiliary field distribution, is identical to
the local field distribution. A sufficient statistic to describe
the field distribution along the dense alignment is $\qbaro$
combined with $q_2$ in the case $\bbar=0$. \cut{, for $\bbar\neq
0$ or $1$ we might anticipate the field distribution to be more
than a combination of these two measures can describe so
additional conjugate parameters may be useful.} The order
parameter $\qbaro$ is related to the correlation of spins,
precisely as the mean spin along the alignment,
\begin{eqnarray}
 \qbaro = \< \frac{1}{N} \sum_i b_i \sigma_i \>\; ;
\end{eqnarray}
this definition is achieved by the derivative of the variational
free energy with respect to $h_D$. Similarly, the variance in the
external fields, $\chi^2$, is conjugate to the parameter $q_2$ in
the replica formalism, and $h_S$ to the mean alignment along
sparse ferromagnetic orientation. $q_2$ (which may also be
determined from the local field distribution (\ref{locfielddist}))
is the SG order parameter, $\qbaro$ and $q_1$ represent
ferromagnetic type order parameters. More generally it is possible
to show that the local field distribution amongst replica spins
(\ref{locfielddist}) is analogous to the field distribution for real
spins in a typical sample from the ensemble $\mI$.

The entropy is an additional physical quantity of interest -- it
is known that this can become negative at low temperatures in
cases where the RS ansatz is insufficient and provides an
indication for an over-simplistic assumption. The energy may be
determined simply from the free energy as
\begin{eqnarray}
\fl e &=& \!-\! \rho/2 \int d\phi(x) x \tanh \beta x \!-\! \Jo\qbaro^2/2 \!-\!
\beta (\Jt q_2^2 \!-\! 1)/2 \!-\! h_D \qbaro \!-\! h_S \int dW(h) \tanh(\beta h)
\nonumber\\
\fl &-&  \rho/2  \int dW(h_1) dW(h_2) d\phi(x) x\frac{(1-\tanh^2(\beta x))
\tanh(\beta h_1) \tanh(\beta h_2)}{1+\tanh(\beta x) \tanh(\beta h_1)
\tanh(\beta h_2)} \!-\! \beta \chi (q_2 \!-\! 1) \label{e}
\end{eqnarray}
and the entropy calculated by the Helmholtz relation
\begin{equation}
 s=\beta(e-\<f\>)\;. \label{s}
\end{equation}
It is possible to see that these thermodynamic extensive
quantities (\ref{s}),(\ref{e}) and (\ref{f}) differ from the
summation of two independent VB and SK subsystems only in the
saddlepoint value of $W(h)$.

\subsection{Longitudinal stability analysis}
\label{LSA}

Recent work has shown that it is also possible to test the so
called longitudinal instabilities within the framework of
population dynamics. Consider that a solution to the equations
(\ref{qbar1rs}),(\ref{q2rs}) and (\ref{Wrs}) can be found, one can
then examine whether the trajectories of fields in the population
$\{h_i\}$ are stable against small perturbations. An unstable
trajectory is indicative of ergodicity breaking and may be
characterised by divergences in certain properties. It has been
shown by Kabashima~\cite{Kabashima:PB} that the method we shall
outline directly tests the longitudinal stability eigenvalue in
the AT formalism~\cite{Almeida:SSK} for the simpler case
restricted to dense couplings. In a sparse coupled system Rivoire
et al~\cite{Rivoire:GM} related the divergence in the mean square
fluctuation to the determination of SG susceptibility through the
fluctuation dissipation relation.

A more general way to test local instabilities in the sparse case
is to consider a less restricted ansatz (1-step of replica
symmetry breaking (1-RSB)~\cite{Mezard:SGT}) on the order
parameters, and determine if the state found is identical to the
simpler ansatz. To our knowledge there is no formalism sufficient
to test all longitudinal stabilities for general topologies. We
would expect our system to conform to the hierarchy of
replica-symmetry broken ansatze, but possibly not be 1-RSB
anywhere. We do not attempt to extend the current analysis to the
1-RSB formalism due to the computational difficulties, especially
at low temperatures; moreover, we find the RS-based analysis to
provide a good description in much of the phase space, and trust
it to be indicative of trends even in the replica-symmetry broken
phases for the simple statistics studied.

Instability within the population dynamics solution arises out of
the microscopic processes occurring in the update equations. In
order to correctly define the consequences of microscopic
variation it is necessary to reformulate the dense part of the
field update ~(\ref{fieldupdateline1}) and represent this by the
microscopic processes of iteration of a full set of fields on a tree (Bethe Lattice), alongside the sparse process. This is equivalent to a reformulation of the problem as belief propagation~\cite{Kabashima:PB}, and testing the convergence of the dynamics to a unique solution.

It is necessary to represent the dense interactions in the
saddlepoint equation~(\ref{eq:saddleq}) by a similar structure to
the sparse interactions, as a summation over many spin-spin
interactions rather than their statistical average. Here, the
saddle point equations of (\ref{eq:saddleq}) take a microscopic
form
\begin{equation}
 b\beta \qbaro + \lambda^{(t)}_1 q_2 \rightarrow \frac{1}{\beta}\sum_j
 b^{(t+1)}_j\atanh\left(\tanh(\beta {\bar x}^{(t)}_j) \tanh h^{(t)}_j \right)\;,
\end{equation}
where ${\bar x}^{(t)}_j$ are now random samples from the dense
coupling distribution (which may be taken as a Gaussian ${\cal
N}(J_0/N,\Jt/N)$), and $b^{(t+1)}_i$ is the
orientation of the spin on the target site, $j$ runs over all
fields excluding those involved in the sparse subsystem. \cut{This
is a microscopic dynamical basis of the saddlepoint equations as
might be established by for example the cavity
method~\cite{Mezard:SGT}, or by analogy with belief propagation.}
We will now calculate how a small variation in the incoming fields
acts to perturb the outgoing field in a particular update.
\begin{eqnarray}
 h_i^{(t+1)} + \delta h^{(t+1)} &=& \sum_{j \setminus \{i,i_1 \ldots i_L\}}
 \frac{b^{(t+1)}_i}{\beta} \atanh\left(\tanh(\beta {\bar x}^{(t)}_j)
 \tanh (h_j^{(t)}+ \delta h_j^{(t)})\right) \nonumber\\
&+& \frac{1}{\beta}\sum_{j=1}^{L} \atanh\left(\tanh(\beta x^{(t)}_{i_j})
\tanh (h^{(t)}_{i_j}+ \delta h^{(t)}_{i_j})\right)
\end{eqnarray}
Where $L$, the set of fields $\{i_j\}$, the sparse and dense
couplings and the target disorder are random samples in any
update. Assuming the perturbations $\delta h$ to be small and
knowing the couplings ${\bar x}$  we can make a Taylor expansion
to linear order obtaining
\begin{equation}
\fl \delta h^{(t+1)} = b_i^{(t+1)}\sum_{j \setminus \{i,i_1 \ldots i_L\}}
(1 - \tanh \beta h_i) {\bar x}^{(t)}_j \delta h^{(t)}_j + \sum_{j=1}^{L}
\frac{(1-\tanh^2(h^{(t)}_{i_j}))\tanh(\beta x^{(t)}_{i_j}) }{1-\tanh(\beta
x^{(t)}_j)
\tanh (h^{(t)}_{i_j})}\delta h^{(t)}_{i_j}\;,
\end{equation}
with the indices again referring to the same quenched samples.
Using the fact that the first part is a sum of $N-L^{(t)} \approx
N$ random variables we can simplify the description of this term
by the central limit theorem, describing the processes by a normal
distribution of mean and variance
\begin{equation}
 \beta \Jo \< (1 - \tanh \beta h_i) \delta h_i \>/N \;;\qquad \beta^2
 \Jt \< (1 - \tanh \beta h_i)^2 \delta h_i \>/N\;,
\end{equation}
respectively.

We are interested in two types of quantities for any solution
\begin{equation}
\chi_1 = \log\left(\frac{\sum \left(\delta h^{(t+1)}_i\right)}{\sum
\left(\delta h^{(t)}_i\right)}\right)^2\qquad;\qquad \chi_2 =
\log\left(\frac{\sum \left(\delta h^{(t+1)}_i\right)^2}{\sum
\left(\delta h^{(t)}_i\right)^2}\right)\;
\end{equation}
which characterise linear and non-linear (SG) susceptibilities by
analogy with results for sparse~\cite{Rivoire:GM} and
dense~\cite{Kabashima:PB} systems. We expect a negative value in
$\chi_1$ to indicate local linear stability, while a negative
value in $\chi_2$ indicates local non-linear stability. We expect
the non-linear stability to be sufficient to detect ergodicity
breaking trends in solutions.

In the case of a paramagnetic solution results may be determined
exactly, the linear instability in $\delta h$ is found to
correspond to the ferromagnetic solution criteria
(\ref{1spinorder}). The non-linear susceptibility is found to
coincide with the SG solution criteria (\ref{2spinorder}).
Therefore the paramagnetic solution instabilities determined in
section~\ref{sec.VB} are reproduced exactly by this method.

Generally we note that by the nature of the population dynamics
algorithm there exist a number of noisy effects: (i) Numerical
finite precision errors; (ii) The noise arising from the random
order in which fields are updated; (iii) the systematic effect of
updating first the field $h^{(t)}_i$, then $ \qbaro^{(t)}$ and
$q^{(t)}_2$; (iv) the finite precision in the histogram; and most
importantly (v) the random sampling of quenched parameters in each
field update. We can therefore expect convergence of our dynamics
to limit not only the final precision achieved, but also the set
of solutions, to those stable against such perturbations.  We
assume this class of perturbations to be sufficient to probe the
linear stability even in the low temperature regime. Any linear
instability should be observable in some moment of the histogram
$\{h^{(t)}_i\}$ so provided we test convergence in $W$ effectively
we hope to rule out linear instabilities.

For the other phases the instabilities depend on the details of
the field distribution. Consider that we have arrived at some
fixed point described by $\{h^{(t)}_i\}$. We can test the
stability of this point by adding small perturbations $\{(\delta
h^{(t)}_i)^2\}$ and determining $\chi_2$ once the dominate modes emerge.
Details of the stability analysis implentation are
given in~\ref{app.PD}.

\section{Numerical results of population dynamics}
\label{sec:results}

We employed population dynamics to investigate a number of systems
concentrating on those in which we could examine the effects of
competing order as both a weak effect (near the high temperature
boundary) and a strong effect at lower temperatures. Many critical
quantities could not be sufficiently resolved at high temperature
 primarily for reasons of finite precision, and near the percolation threshold, in our
algorithms we present cases with ($\beta/\beta_c<10$) and $\rho=2$, $1/\beta_c$ being the largest critical temperature (typically $1$ for examples chosen).

\subsection{Data format}
All data is based on linear spline interpolation of array data. In
the figures over the range $\gamma=(0,1)$ we use a point spacing
of $(\delta \beta,\delta \gamma)=(0.025,0.025)$. For these
diagrams we present results based on a single run from random
initial conditions, and plot the mean and error bars (linear spline interpolation
based on values at each sample point) over $20$
samples of data. These $20$ samples are selected in successive
time steps (a time step is $N$ single field updates, $1$ for each
field), following the convergence of the distribution. Therefore,
the error bars are not over independent samples, and as such not
necessarily well described by an uncorrelated Gaussian
distribution, but this assumption gives a first approximation to
single time-step fluctuations in the measured state. Each point
that did not converge within $500$ time steps is marked by a
cross.

Figures that focus on part of the $\gamma$ range use point spacing
of $(0.005,0.005)$. We average quantities in these figures over 10
runs based on different seeds for the random number generator
(different initial conditions and updates). The data point is
taken to be the mean of $20$ samples from each run. Unlike the
sampling within a single run (which may be correlated in
successive update steps) we can be confident of the independence
of these samples conditional on the parameterisation and
convergence criteria, and present the mean and error bars for
quantities of interest (again interpolation).

Population dynamics is not required in the paramagnetic region,
for which simple boundary conditions based on exact knowledge are
imposed, this creates some unevenness in certain quantities very
close to the boundary. A second source of unevenness is the array
like nature of the data points. Finally our convergence criteria appears not
strict enough to prevent some systematic errors (drift) near critical regions,
otherwise we expect error bars to be representative.

Cases of incomplete convergence are treated
for purposes of data collection and interpolation as converged results. Incomplete convergence
occurs:(i) in a small number of cases close to critical transitions, (ii) in the case of competing
ferromagnetic alignment. After the maximum number of population
iterations is completed ($500$) we select the histogram (amongst 4
differing in initial conditions) of minimum free energy which is
good in case (ii), but does not resolve case (i). However, we
found experimentally that in cases (i) many quantities did not
vary greatly in absolute terms between the histograms of different
initial conditions, except for some systematic drift in the stability parameter $\chi_2$.

\subsection{Figures presented and coupling distributions considered}
As there is a large number of composite systems that could be
considered and analysed, it would be useful to review the choice
of the specific cases presented here.

We present data for the several aligned ($\bbar=1$) combinations of sparse and
dense subsystems. A mixture of a dense subsystem of ferromagnetic
couplings with a sparse SG (figure~\ref{figFSG}) or
anti-ferromagnetic (figure~\ref{figFAF}) couplings; and the
converse, a subsystem of sparse ferromagnetic couplings mixed with
a dense subsystem of SG (figure~\ref{figSGF}) or anti-ferromagnetic
(figures~\ref{figAFF} and~\ref{figAFFdisc_cont}) couplings. We
also consider the case of two subsystems both with ferromagnetic
couplings but unaligned ($\bbar=0$, figures~\ref{figFF0} and ~\ref{figFF0inset}).

The dense couplings are parameterised by $(J_0,\Jt)$, so that for example a dense anti-ferromagnetic
system is $(-1,0)$ in this notation. The sparse system we choose
is the $\pm J$ model, a canonical case
convenient for numerical reasons. In this model $\phi$ is bimodal
and described by $p$ the probability of sampling $+J$ as opposed
to $-J$ for each bond. Therefore the sparse part is parameterised by $(p,J)$ with $\rho=2$,
describing the relevant subsystems we present.

Owing to the large variety of parameterisations we are not able to
test sufficient systems to be able to make so general statements
as are possible in the exact analysis of the high temperature
behaviour. However, we believe these systems to be the most
intuitive of combinations. We examined the effect of changing the
distribution of sparse couplings $\phi$ from $\pm J$ to Gaussian
but found only marginal variations.

\subsection{Stability Results}

Results are \emph{generally} characterised by a dominance of the
dense system below a critical mixing parameter $\gamma_c$ which is
replaced by a dominance of the sparse system at larger $\gamma$, although
the critical value and the specific properties depend on the
system studied and the temperature value.

We find that everywhere close to the boundary the longitudinal
stability eigenvalue is positive, thus the RS solutions appear to
be locally stable, including in the SG phase. The small gap observed for
the SG solution between the numerical results obtained from
population dynamics and the high temperature transition line is we suspect due to
finite size effects in a combination of population size and number of updates.
Otherwise the SG solutions are unstable
to ergodicity breaking as expected. The RS ferromagnetic solution
may also become unstable at sufficiently low temperatures.
Therefore it appears that there is a mixed phase as predicted by the stability analysis in which the
magnetic moment is non-zero but ergodicity breaking is present.
Dense ferromagnetic phases appear less susceptible to this instability.

\subsection{The Co-emergence regime: properties on lowering temperature about a
P-F-SG triple-point,
$\gamma\approx\gamma_C$} We can examine the type of order observed
as we decrease the temperature from the triple-point, which
provides information as to the relative importance of couplings in
weakly ordered systems. It appears that in the vicinity of the
triple point one can lower the temperature and finds that the type
of order present is closer to that induced by the dense couplings
in most cases (inline with the predictions of the high temperature
analysis). The sparse type order parameter in some cases completely
disappears as temperature is lowered.

If the dense type order is ferromagnetic (figure \ref{figFSG}) it
appears that by lowering temperature a region of ergodicity
breaking may be encountered, but the magnetic moment does not
disappear; such a scenario describes a mixed phase. Conversely if
one lowers the temperature where the high temperature is a
sparsely induced ferromagnetic one (figure \ref{figSGF}), it is
possible not only for the ferromagnet to become unstable towards a
mixed phase, but for the magnetic moment to disappear entirely.
This characterises a standard reentrant behaviour, as the magnetic
moment re-enters the value it had in the paramagnetic phase. For
the anti-ferromagnetic SG state (figure \ref{figAFF}) there are some
indications of transitions from SG to mixed as well as
ferromagnetic to mixed transitions as temperature decreases.

A more exotic possibility observed in the simple combination of a
dense ferromagnet and sparse SG subsystems (figures~\ref{figFSG}
and \ref{figFAF}) is that a SG phase may become an RS ferromagnet
and then a mixed phase as temperature is lowered. The numerical
results are inconclusive on this point, but one can combine this
with knowledge of the exact transition line result at the boundary
(\ref{criticalline}), and the intuition that ergodicity breaking
is more likely at low temperatures in frustrated systems. It is
unusual to our knowledge for ergodicity breaking to disappear as
temperature is lowered, only for it then to reappear.

\subsection{Discontinuous high temperature transitions}
\label{sec:HTT}
\begin{figure*}[htb]
\begin{center}
\scalebox{1}{\includegraphics*{{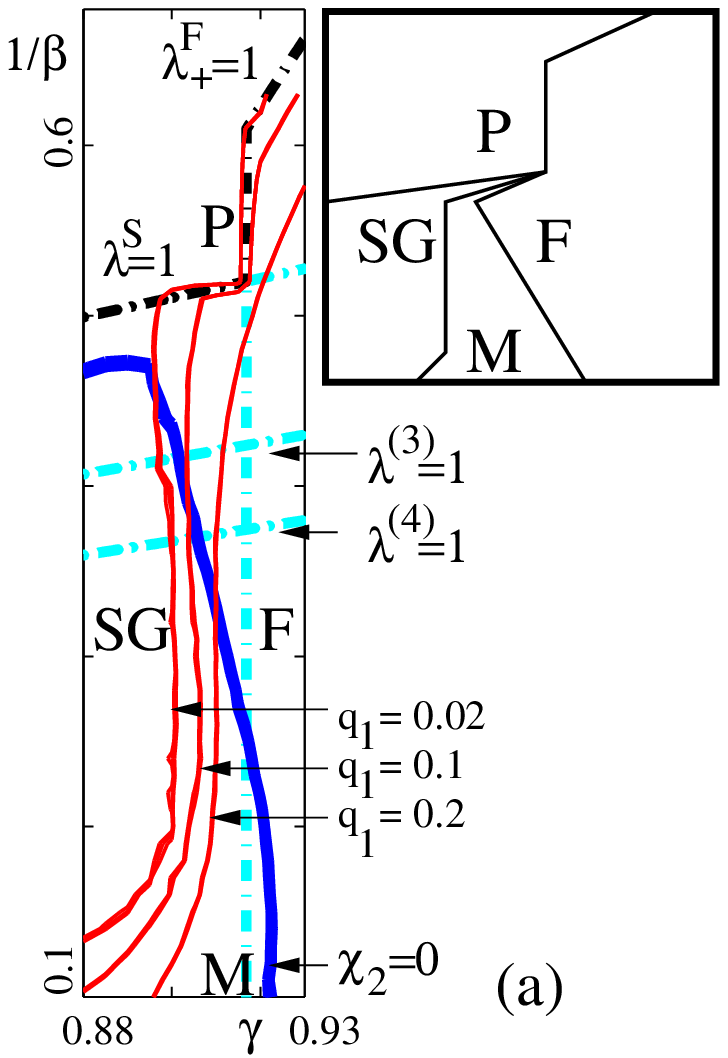}}}
\scalebox{1}{\includegraphics*{{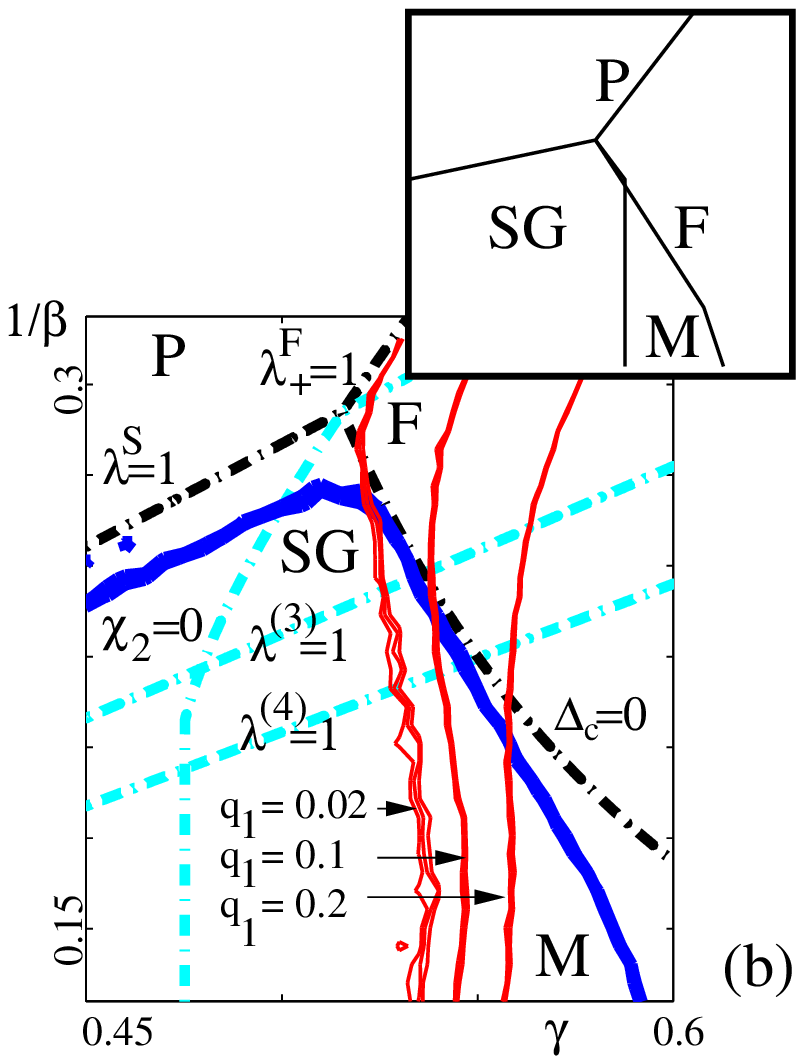}}}
\caption{\label{figAFFdisc_cont} Anti-ferromagnetic couplings in the dense part aligned with ferromagnetic order in sparse couplings can cause the high temperature ferromagnetic solution criteria (\ref{1spinorder1}) to be unmet for any $\beta$ in some range of $\gamma$. If $\ref{1spinorder}$ describes the high temperature transition as $\gamma$ decreases towards this range then there is a discontinuity (a) and the stability analysis for the region $\Delta_2=0(\Delta_1)$ does not apply, otherwise the transition is a continuous one (b) with the stability analysis applicable. 
(a) With antiferromagnetic dense couplings $(\Jo=-2,\Jt=0)$ and ferromagnetic sparse couplings $(p=1,J=1)$, the transition is discontinuous. The sparse order dominates the dense order near $\gamma\approx\gamma_C$, the
relative importance of the dense couplings increases
as temperature is lowered. The closeness of the magnetisation contours to the
line $\lambda^S=1$ may be a significant finite size effect, these lines do not coincide in the thermodynamic limit.
(b) With antiferromagnetic dense couplings $(\Jo=-\frac{1}{4},\Jt=0)$ and ferromagnetic sparse couplings $(p=1,J=1)$, the transition is continuous. Results appear to follow closely the high temperature prediction. At lower temperature $\gamma \approx \gamma_C$ appears to
remain approximately coincident with a SG-M transition, the ferromagnet is again unstable to a mixed phase in many systems.
}
\end{center}
\end{figure*}

Figures ~\ref{figAFF} and~\ref{figAFFdisc_cont} consider
scenarios in which the high temperature transition is potentially
\emph{discontinuous}. Where the boundary is discontinuous, the
sparse coupling order is ferromagnetic; it appears that the
sparse-induced ferromagnetic order overwhelms any emergence of SG
order in the vicinity of the discontinuity ($\gamma \approx
\gamma_C$), so that the SG phase must occupy only some very narrow
region about the high temperature transition. We can anticipate
for both the SG and paramagnetic phases, that convergence of
population dynamics might be subject to especially strong finite
size effects in the vicinity of this transition. We vary the
sparse ensemble through $J$ to generate the different scenarios, $\rho$
may also be varied to create this effect.

\begin{figure*}[htb]
\begin{center}

\includegraphics*{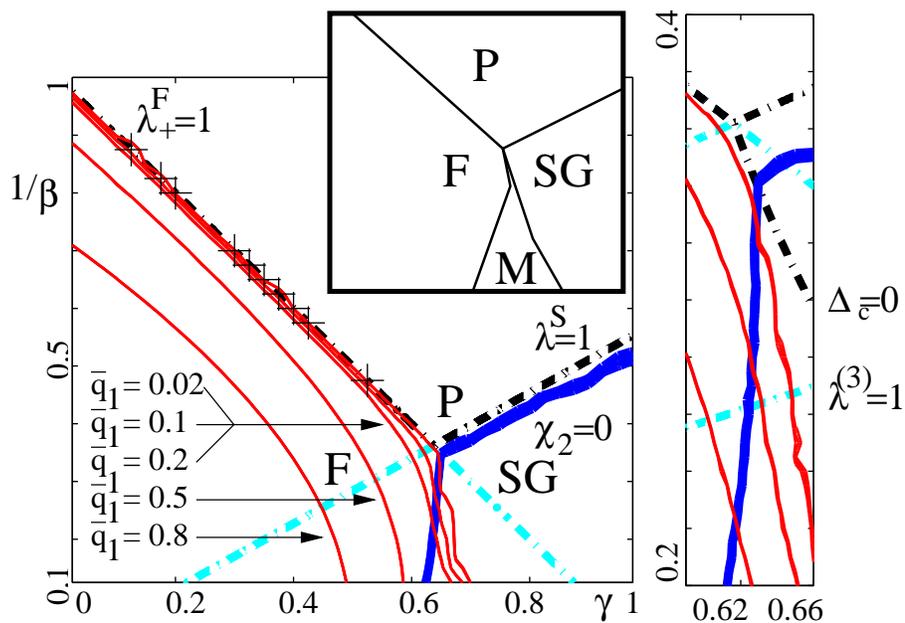}
\caption{\label{figFSG} A dense subsystem $(J_0=1,\Jt=0)$ with a
sparse subsystem $(p=0.5,J=1)$ -- dense ferromagnetic and
sparse SG couplings. A schematic denotes our qualitative
understanding of the solutions. Main figure: The exact critical
lines ((\ref{2spinorder}),(\ref{pspinorder}),(\ref{1spinorder}))
are indicated by dot dashed lines with the darker (upper) line
being the exact high temperature transition. The thick line
$\chi_2=0$, and the line of zero magnetisation
$q_1=0$, indicate transition points as determined by the population
dynamics algorithm. The continuation of $q_1=0$ must be inferred
by the contours in $q_1$ (thin lines). Error bars in all contour lines are
plotted and indicate time-step fluctuations within a single run.
Transitions are between unstable-stable and SG-F solutions, the
unstable ferromagnetic solution is classified a mixed state. It
appears the effect of sparse frustrating bonds is to make the RS
dense ferromagnet unstable as $\beta$ increases. Similarly the
sparsely induced SG may be susceptible to a transition towards a
mixed state with increasing $\beta$. The mixed phase remains close
to $\gamma=\gamma_c$, thus the competition of interaction types
remains in some sense balanced as temperature decreases. Inset
figure: The right figure shows the same contours as the main figure in
magnification with corresponding error bars (variation between
different independent minimisations). We also plot the high
temperature critical line (\ref{criticalline}), which indicates
the exact gradient for the curves $\chi_2=0$ and $q_1^2=0$, this
appears not to be closely followed by the line $\chi_2$ which is
due to finite size numerical errors and coarseness of point
sampling.}
\end{center}
\end{figure*}

\begin{figure*}[htb]
\begin{center}
\includegraphics*{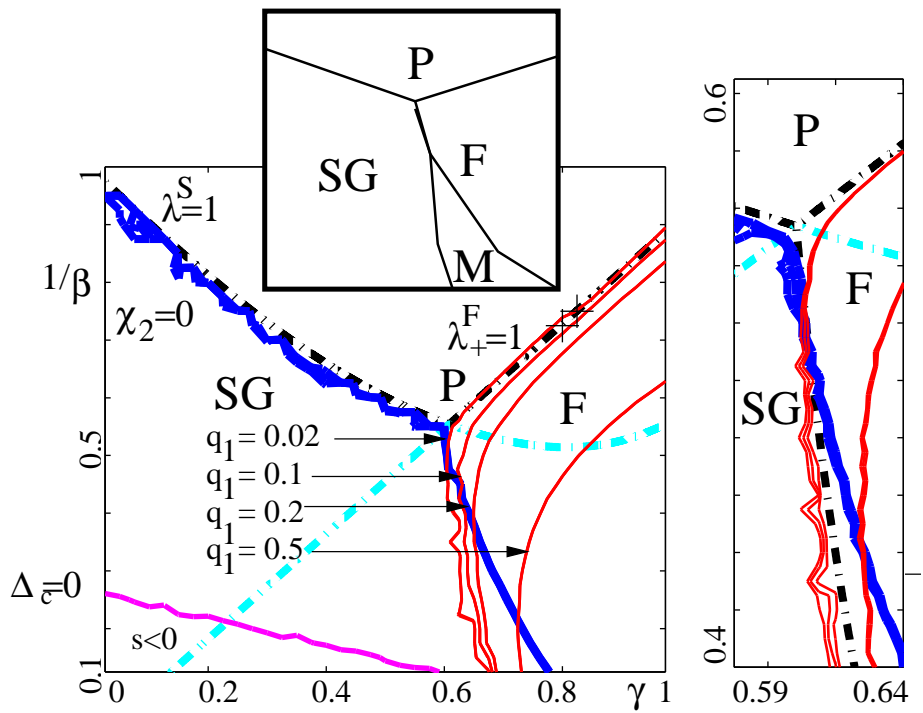}
\caption{\label{figSGF} A dense subsystem $(J_0=0,\Jt=1)$ with a
sparse subsystem $(p=1,J=\frac{1}{2})$, i.e., dense SG and sparse uniform
ferromagnetic couplings. A schematic denotes our qualitative
understanding of the solutions. Main figure: Lines and symbols are
the same as for figure~\ref{figFSG}, in addition the thick dotted
line represents a lower temperature bound to positive entropy
(include the negligible time-step errors), below this line the
method is inconsistent as would be suggested by our results on
instability of the SG phase (to which this curve belongs). Inset:
The low magnetisation contours, and stability parameter follow
quite closely the triple-point analysis prediction
(\ref{criticalline}) in the vicinity of the boundary (inset).
These lines break indicating a mixed phase. It appears there may
be no transitions in the SG state towards, a mixed phase. The
mixed phase is shifted to higher $\gamma$ with increasing $\beta$
indicating the greater relative importance of the dense disordered
couplings in determining phase behaviour as temperature is
lowered. The ferromagnetic state is first unstable to a mixed
state, which may in turn be unstable towards a SG state.}
\end{center}
\end{figure*}

\begin{figure*}[htb]
 \begin{center}
\includegraphics*{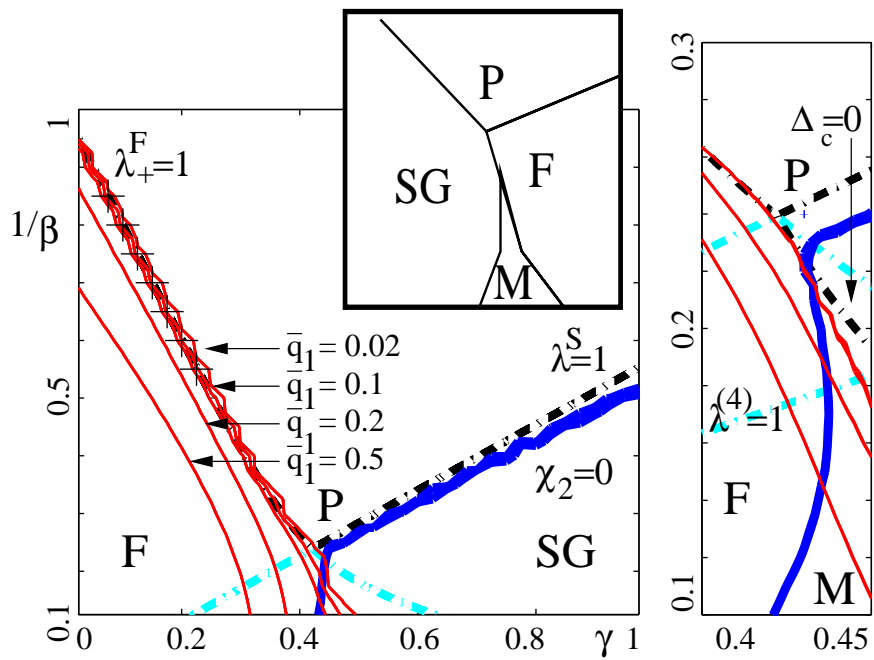}
\caption{\label{figFAF} The combination of dense subsystem $(J_0=1,\Jt=0)$
and sparse subsystem $(p=0,J=\frac{1}{2})$ is a combination of ferromagnetic and
antiferromagnetic (SG) effects; lines and symbols are as in the
previous figures. Not surprisingly this case is almost identical
to the case of figure~\ref{figFSG}. There is excellent agreement
between the thin data lines (ferromagnetic moments) and the high
temperature prediction for their disappearance (inset). A clear
cusp exists in the stability, indicating that an ergodic
ferromagnetic behaviour may emerge in some systems as temperature
is lowered, this indicates increasing importance of the dense
couplings as temperature is lowered. However, instabilities do
reemerge at lower temperature. The SG is susceptible to a mixed
state as temperature is lowered.}
\end{center}
\end{figure*}

\begin{figure*}[htb]
\begin{center}

\includegraphics*{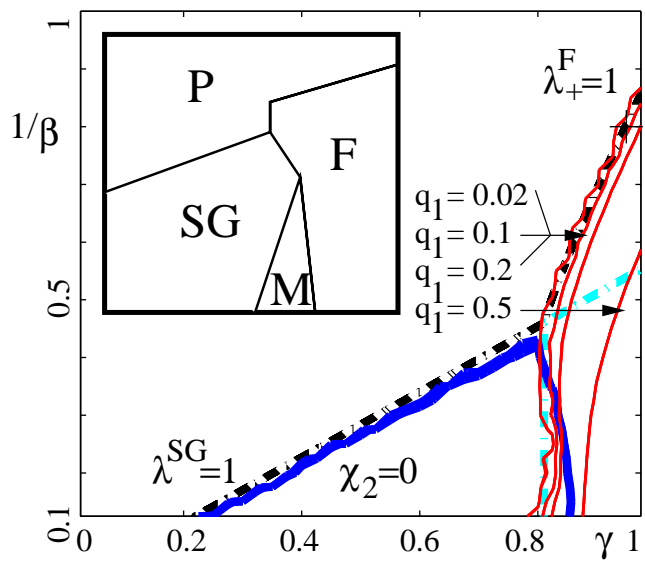}
\caption{\label{figAFF} The combination of dense subsystem
$(J_0=-1,\Jt=0)$ and sparse subsystem $(p=1,J=\frac{1}{2})$ is a combination of
antiferromagnetic and ferromagnetic effects; lines and symbols as
in the previous figures. The dense couplings cannot independently
induce SG or ferromagnetic order, but do suppress ferromagnetic
order. The resulting behaviour is comparable to a SG ferromagnet
combination-- figure~\ref{figSGF}. By contrast to the dense SG
coupling there is weaker suppression of the magnetic moment at low
temperature, and ergodicity breaking is a weaker effect at low
temperature in the ferromagnetic phase. In this diagram the high
temperature transition line are marginally discontinuous as
discussed (section~\ref{sec:HTT}). Therefore we do not provide a
detailed inset for the triple point analysis region. }
\end{center}
\end{figure*}

With the transition nearly discontinuous (as is the case for
$J=1/2$ figure~\ref{figAFF} in which the discontinuity is visually
imperceptable) one observes a
similar behaviour to the case of a SG-F combination
(figure~\ref{figFSG}) except in the weaker suppression of the
ferromagnetic order at low temperatures. Clearly, in this case the
antiferromagnetic dense component cannot induce any order except
for a paramagnetic one and the emergence of a SG is a property of
the sparse distribution.

In figure \ref{figAFFdisc_cont}a where a discontinuity is clearly
visible we find ergodicity breaking is surpressed in the vicinity
of the discontinuity. The ferromagnetic phase dominates near
$\gamma_c$ at high temperature but gives way to a mixed phase at
lower temperature. Figure ~\ref{figAFFdisc_cont}b shows that in the
continuous case the high temperature prediction is qualitatively
accurate. The SG state appears to dominate near $\gamma_c$, with
mixed phases appearing at lower temperature.

\subsection{Unaligned ferromagnetic couplings $\bbar=0$}
\label{ssec:UFC}
\begin{figure*}[htb]
\begin{center}

\includegraphics*{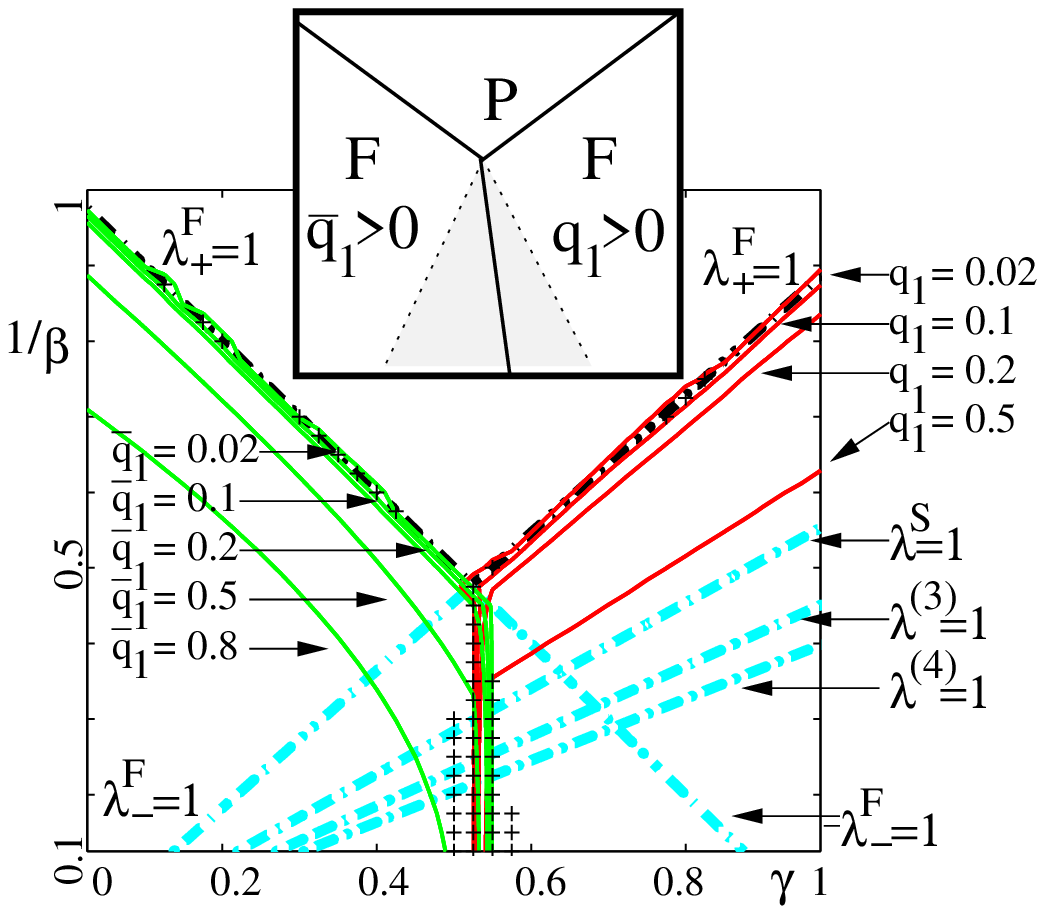}
\caption{\label{figFF0} A system with $\bbar=1$, and
uniform couplings of $(J_0=1,\Jt=0)$ and $(p=\frac{1}{2},J=1)$, models
the competition between two unaligned ferromagnetic orderings;
lines and symbols as in the previous figures. The converged
solutions were found to be ergodic everywhere ($\chi_2<0$) and
were of positive entropy. In a part
of the space below the cusp in the high temperature transition
($\gamma_c$) population dynamics converge to one of two differently aligned locally stable states as can be seen by the crosses.
Varying $\gamma$ through the centre of this region there is a rapid decline in $\qbaro$ with the magnetic moment growing instead in $q_1$. We speculate that this is a first order transition between two locally stable states.}
\end{center}

\end{figure*}

This is an interesting case of competing ferromagnetic alignments;
as one decreases the temperature in the vicinity of $\gamma_C$ it
appears that, as a thermodynamic solution, the sparse alignment is marginally
dominated in the thermodynamic state by the dense alignment (figure \ref{figFF0}). However, the sparse aligned
solution persists at lower temperature as a metastable state,
responsible for the non-convergence, which is locally stable
against ergodicity breaking ($\chi_2<0$).

\begin{figure*}[htb]
\begin{center}

\includegraphics*{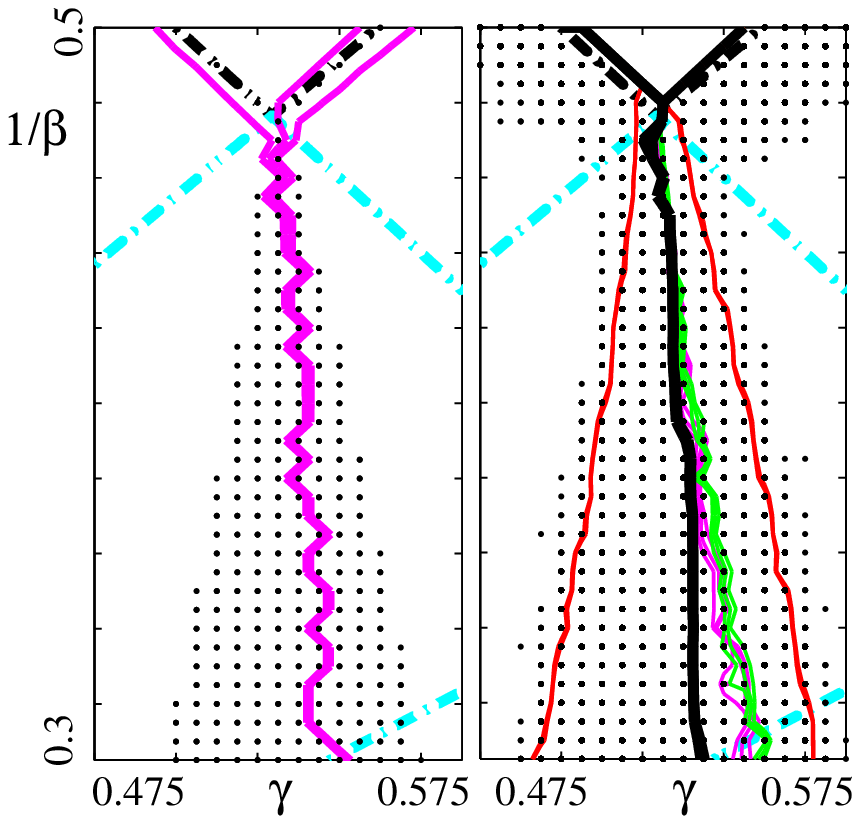}
\caption{\label{figFF0inset} A system with $\bbar=1$, and uniform
couplings of $(J_0=1,\Jt=0)$ and $(p=\frac{1}{2},J=1)$, models the
competition between two unaligned ferromagnetic orderings. These
figures show magnifications of figure \ref{figFF0} near
$\gamma_c$. Left figure: In the vicinity of $\gamma_c$ a large
number of states fail to converge after maximum algorithm time of
$500$ steps (dots). We then have to chose which of the 4
replica histograms evolved in the dynamics produces a result of
smallest free energy. If we evaluate the magnetisation in this
state we find a transition indicated by change of sign in
$\qbaro-q_1$, this is plotted with error bars (time step
fluctuations) for a single run. Left figure: Sampling 20 systems
with a maximum time of $100$ time steps in each iteration we show
the contour $\qbaro-q_1=0$ with sampling errors for several
different histograms evolved in parallel: a sparsely aligned
ferromagnetic (left most line), random SG or nearly paramagnetic
(central thin lines) or densely aligned ferromagnetic initial
histogram. The central thicker black line is the equilibrium
result as determined by from the histogram of minimum free
energy. With only $100$ iterations there is a slightly broader set
of unconverged states (dots), but we already see the emergence
of two well defined dynamical transitions - left and right lines,
matching approximately the result in the left figure. The
alignments of the random histograms appear to approximately match
the thermodynamical transition at high temperature but are noisy
and slightly biased towards a dense alignment at lower
temperature.}
\end{center}
\end{figure*}

In fact, metastable states of both dense and sparse alignment
exist across a wide range of parameters $\gamma$ for any $\beta$
below the triple point (figure \ref{figFF0inset}), indicated by populations converging
to different numerically stable solutions. These metastable states
are characterised numerically by $\qbaro>0,q_1\approx 0$ and $q_1>0,\qbaro\approx0$ and we
speculate the thermodynamic state involves a first order transition between the two solutions.
A necessary condition for the existence of
a metastable state is the non-negativity of both
$\lambda^F_\pm-1$, but this is not necessarily sufficient.

Histograms with a well defined alignment tend to converge to the locally stable state with the same corresponding alignment, whether or not this is the thermodynamic solution. Random initial conditions appear to converge towards the thermodynamic solution at high temperature, but have a slight bias towards the dense alignmnent at lower temperatures (atleast over a small number of updates).
It is difficult to determine if the solution space generate a larger basin of attraction for
the dense metastable state, or one, in some sense,
proportional to the free energy. This may be important given that
we envisage composite systems as components in optimisation
problems were dynamical minimisation by local search, similar to
population dynamics, is a critical feature, possibly more so than
any static properties of the ensemble.

\section{Conclusion}
\label{sec:conclusions}
This paper considers the question of systems consisting of a
combination of sparse and dense random couplings. Given the
complementary properties of these types of structures, combined
with the possibility of a robust statistical description, many
engineered systems may benefit by a decomposition as laid out
here. Such an example we are currently pursuing involves the
combination of sparse and dense spreading codes for multiuser
access~\cite{CDMA_RS}. An alternative application of this analysis
might be in modelling random attacks on network structures, where
correlations between attacked elements are induced~\cite{Hase}.

This paper has considered a simple model in which canonical sparse
and dense disordered models have been mixed. In so doing we have
observed some clear trends which may generalise. These include the
tendency for dense induced order to dominate sparse induced order
when in competition. In a system with a dense ferromagnetic
tendency, it is possible for this order to emerge from a sparse
induced SG as a replica-symmetric stable system by lowering the
temperature, provided the order is sufficiently weak (high
temperature). In the vicinity of a high temperature transition the
effect of suppressing ferromagnetic order by an increasing
antiferromagnetic tendency in the bonds may have a very similar
effect to the introduction of frustrating effects in the
interaction structure. Finally, the case of the unaligned system
indicates that the dominance of dense ferromagnetic  over sparse
ferromagnetic couplings is marginal in equilibrium properties,
however it appears a metastable behaviours may have non-trivial consequences for dynamics.

Several questions remain open - importantly, the question of the
low temperature limit is not resolved in this paper nor is the behaviour in the vicinity of the percolation threshold. These may be analysed by related statistical physics methods. There are a
number of other variations which would allow the insight gained
from the current study to be strengthened, including
reformulations of the basic model so that the arbitrary parameter
$\gamma$ may be shown to be only a technicality; so that related
models, for example, in which a dense system is fixed and one
gradually adds more bonds (a variation of $\rho$), may be
surmised. One can consider a number of ways of scaling
$\phi$,$J_0$ and $\Jt$ with $\gamma$ but we hope the properties we
have reported here are robust against the most sensible
alternatives such as scaling all coupling moments of each
subsystem uniformly with $\gamma$.

Finally, with the increasing interest in complex systems of
varying connectivities and interaction strengths, we believe the
current study of a simple composite system comprising elements
which have both sparse (and strong) and dense (weak) interactions,
represents a first step in the principled analysis of their
equilibrium behaviour.

\subsubsection*{Acknowledgments}
DS would like to thank Ido Kanter for useful comments. This work
is partially supported by EVERGROW, IP No. 1935 in the complex
systems initiative of the FET directorate of the IST Priority, EU
FP6 and EPSRC grant EP/E049516/1.

\clearpage

\appendix
\section{Replica calculation}
\label{app.RC}
A particular ensemble $\mI$ is described by a set of order one scalars
$\{J_0,\Jt,\rho,h_S,h_D,\chi^2\}$ and a sparse coupling distribution $\phi$ of
finite moments on the real line, without finite measure at zero. These parameters contain an
unstated dependence on the mixing parameter, we choose this to be in the
couplings $\phi$,$J_0$ and $\Jt$ but leave this unstated for brevity. An
instance of the ensemble is constructed according to the following probability
distributions, subscripts being the relevant random variables
\begin{eqnarray}
 P_\vb(\vx) &\propto& \prod_{i=1}^N \left[\delta_{x_i,1}(1+\bbar) +
\delta_{x_i,-1}(1-\bbar)\right]\;;\nonumber\\
P_{\matJ^D}(\matx) &=& \prod_{\ij} \phi^D(x_\ij) \;;\nonumber\\
P_{\matJ^S}(\matx) &=& \prod_{\ij} \phi(x_\ij) \;;\nonumber\\
P_{\matA}(\matx) &=& \prod_{\ij}
\left[\left(1+\frac{\rho}{N}\right)\delta_{x_\ij}
+ \frac{\rho}{N} \delta_{x_\ij,1} \right] \;;\nonumber\\
P_{\vh}(\vx) &=& \prod_i \phi^F(x_i) \;.\nonumber
\end{eqnarray}
The dense coupling distribution $\phi^D$ may be taken as a Gaussian of mean
$J_0/N$ and variance $\Jt/N$. The field distribution for each site is a Gaussian
of mean $h_D b_i+h_S$ and variance $\chi^2$.

We describe the properties of a particular ensemble through self averaged
quantities calculated from the mean free energy density (self averaging
assumption) which with the replica identity~\cite{Mezard:SGT} may be written
\begin{equation*}
 \<f\> = \lim_{N\rightarrow \infty} - \frac{1}{\beta N} \lim_{n \rightarrow 0}
\frac{\partial}{\partial n} \< Z^n \>\;,
\end{equation*}
with $\beta$ the inverse temperature. The replicated partition function,
averaged over samples is given by
\begin{eqnarray}
\fl \<Z\> &=& \sum_{\vS_1 \ldots \vS_n} \< \<\exp\{ \sum_{\alpha} \sum_\ij b_i
b_j J^D_\ij \sigma^\alpha_i \sigma^\alpha_j \}\>_{\matJ^D} \<\exp\{
\sum_{\alpha} \sum_\ij A_\ij J^S_\ij \sigma^\alpha_i
\sigma^\alpha_j\}\>_{\matA,\matJ^S} \right. \nonumber \\
\fl &\times& \left. \<\exp\{ \sum_{\alpha} \sum_{i} h_i \sigma^\alpha_i
\}\>_{\vh} \>_{\vb}\;. \nonumber
\end{eqnarray}
As is standard in such calculations~\cite{Mezard:SGT} we maintain only those
terms in the exponent which are order $N$, taking asymptotic values in
$n\rightarrow 0$ for brevity in all prefactors.
The quenched average on dense couplings may be factorised and taken by
linearising the exponent, resulting in
\begin{equation*}
 \< \cdots \>_{\matJ^D} = \prod_\ij \exp \{\frac{\beta \Jo}{N} b'_i b'_j
\sum_\alpha \sigma^\alpha_i\sigma^\alpha_j + \frac{\beta^2 \Jt}{2 N}
\left(\sum_{\alpha}  \sigma^{\alpha}_i\sigma^{\alpha}_j \right)^2\}\;.
\end{equation*}
The sparse coupling average may also be taken by standard sparse
methods~\cite{Monasson:OP}
\begin{equation*}
 \fl \< \cdots \>_{\matJ^S,\matA} = \prod_\ij \left[\exp\{-\frac{\rho}{N}\}
\exp\{\frac{\rho}{N}\left(\< \exp \{\beta x \sum_\alpha \sigma^\alpha_i
\sigma^\alpha_j\}\>_{\phi(x)}\} - 1 \right)  \right]\;,
\end{equation*}
and the field average
\begin{equation*}
\< \ldots \>_{\vh} = \prod_i \exp \{\beta (h^D b_i + h^S)\sum_\alpha
\sigma^\alpha_i +  \beta^2\chi^2/2(\sum_\alpha \sigma^\alpha_i)^2 \}\;.
\nonumber
\end{equation*}
Introducing the following order parameters, $\Phi$:
\begin{equation*}
\fl \pi(\vsigma) = \frac{1}{N}\sum_i \delta_{\vsigma_i,\vsigma} \;; \nonumber \qquad
\qbaral = \frac{1}{N}\sum_i b_i \sigma^\alpha_i \;;\qquad \qalal = \frac{1}{N}\sum_i \sigma^{\alpha_1}_i \sigma^{\alpha_2}_i\;; \label{orderparamdefs}
\end{equation*}
we are able to write the replicated partition function as an integral in the set of order parameters
\mathindent =0pc
\begin{eqnarray}
 \< Z^n\> &\propto& \int  d\Phi \sum_{\vsigma_1 \ldots \vsigma_N} \fatI(\Phi)
\exp\{-N G_1\} \;;\nonumber\\
 \phantom{G} G_1 &=& -\frac{\beta J_0}{2}\sum_\alpha \qbaral^2 -
\frac{\beta^2 \Jt}{2}\sum_\alpha \qalal^2 -
\frac{\rho}{2}\sum_{\vsigma_1,\vsigma_2} \pi(\vsigma)\pi(\vsigma)
\<\exp\{\beta x \sigma^{\alpha_1}
\sigma^{\alpha_2}\}-1\>_\phi \nonumber\\
 &-& \beta h_S \sum_\vsigma \pi(\vsigma) \sigma_\alpha - \beta
h_D \sum_\alpha q_\alpha - \beta^2 \chi^2\sum_\alal \qalal
\;;\nonumber
\end{eqnarray}
where the function $\fatI$ is an indicator function for the order parameter
definitions ~\ref{orderparamdefs}. We note here that there is potentially some redundancy in the definition of
order parameters, this is allowed for a concise and general expression.

Representing each definition in the indicator function by a Fourier transform,
introducing conjugate integration variables denoted with a hat,
\begin{eqnarray}
\sum_{\vsigma_1 \ldots \vsigma_N} \fatI(\Phi) &=& \int d\Phi \exp \{-N (G_2
+ G_3)\}\nonumber\;;\\
 G_2 &=& -\sum_\alpha \qbaral \qhbaral -\sum_\alal \qhalal \qalal - \sum_\vsigma \nonumber
\pi(\vsigma)\pih(\vsigma) \nonumber\;;\\
 G_3~&=& -\log \sum_\vsigma \Avb{\exp\{ - b \sum_\alpha \qhbaral \sigma^\alpha - \sum_{\alal} \qhalal \sigma^{\alpha_1} \sigma^{\alpha_2}  - \pih(\vsigma) \} }\nonumber\;;
\end{eqnarray}
assuming $N$ is the suitable scaling for the conjugate variables. We have
finally a site factorised saddlepoint form (\ref{basicsaddlepointform}).

\section{Calculation details for auxiliary system}
\label{app.VB}

In order to construct the auxiliary system we must determine the
significance of terms. We keep terms sufficient to allow a second
order description in $\Delta$, which is the perturbation away from
$\lambda_S=1$ and/or $\lambda^F_+=1$, and to allow a second order
stability analysis, restricted to the region where $\lambda^{(p)}<1$ and
$\lambda^{F}_-<1$. We calculate the expansion of the complicated entropic term 
$\log\Tr\exp\{\mX\}$ in equation~(\ref{Lambda2}) in the reduced
set of order parameters $\{\tal,\Qalal,\Qalalal,\Qalalalal\}$ and
to an order fit for purpose. 
This becomes for $\bbar=1$ ($v_2=1$) or $\bbar=0$ ($v_2 =1$ or $0$)
\begin{eqnarray}
- \fl\log \Tr \exp \mX &=&\! -\sum \log \cosh \tal \!-\! \sum \log
\cosh \Qoalal \label{mlogTrexpmX}\\
&-& \sum \log \cosh  \Qoalalal^2
-\sum \log \cosh \Qoalalalal^2 \nonumber  \\
&-& \log \Tr
\<\prod \left[1 + v_2 \tanh(\tal) \sigma_\alpha \right]\>
\prod \left[ 1 +
\tanh(\Qoalal) \prod_{i=1}^2 \sigma_{\alpha_i} \right] \nonumber\\
&\times&  \prod \left[ 1 +
\tanh(\Qoalalal)\prod_{i=1}^3\sigma_{\alpha_i}\right] \prod \left[
1 + \tanh(\Qoalalalal) \prod_{i=1}^4\sigma_{\alpha_i}\right] \nonumber\ ;
\end{eqnarray}
the sums and products are ordered in the various terms and are
over the corresponding replica indices. The case of an aligned
system and unaligned have simple averages, being distinguished by $v_2$. The Trace requires a graphical expansion, otherwise various non-linear terms must in some cases be expanded upto third order. The expansion of (\ref{mlogTrexpmX}) then gives
\begin{eqnarray}
&-& \frac{1}{2} \sum \tal^2 - \frac{1}{2! 2} \sum
Q_{\alpha_1 \alpha_2}^2 - \frac{1}{ 3! 2}\sum \Qalalal^2 - \frac{1}{ 4! 2} \sum \Qalalalal^2 \nonumber\\
\fl &+& \frac{1}{12}\sum \tal^4 + \frac{1}{2!12 }\sum Q_{\alpha_1 \alpha_2}^4 - \frac{1}{45}\sum \tal^6 \nonumber \label{Lambdaapp}\\
\fl &-& \frac{1}{2!}\sum t_{\alpha_1}t_{\alpha_2} Q_{\alpha_1 \alpha_2} \!+\! \frac{2}{2! 3 }\sum t_{\alpha_1}^3 t_{\alpha_2} Q_{\alpha_1 \alpha_2} \!+\! \frac{1}{2! 3 }\sum t_{\alpha_1} t_{\alpha_2} Q_{\alpha_1 \alpha_2}^3 \!-\! \frac{3}{3!}\sum t_{\alpha_1} t_{\alpha_2} Q_{\alpha_2 \alpha_3} Q_{\alpha_1 \alpha_3} \nonumber\\
\fl &-& \frac{1}{3!} \sum Q_{\alpha_1 \alpha_2} Q_{\alpha_2 \alpha_3} Q_{\alpha_1 \alpha_3} \!-\! \frac{12}{4!}\sum t_{\alpha_1} t_{\alpha_2} Q_{\alpha_1 \alpha_3} Q_{\alpha_2 \alpha_4} Q_{\alpha_3 \alpha_4} \!-\! \frac{3}{4!} \sum Q_{\alpha_1 \alpha_2} Q_{\alpha_1 \alpha_3} Q_{\alpha_2 \alpha_4} Q_{\alpha_3 \alpha_4}\nonumber\\
\fl &-&  \frac{3 v_2}{3!}\sum t_{\alpha_1} Q_{\alpha_2 \alpha_3}\Qalalal \!-\! \frac{v_2}{3!}\sum t_{\alpha_1} t_{\alpha_2} t_{\alpha_3} \Qalalal - \frac{3}{4!}\sum \Qalalalal Q_{\alpha_1 \alpha_2} Q_{\alpha_3 \alpha_4} \nonumber
\end{eqnarray}
where sums are now over all sets of indices without repeated
indices. Finally introducing the energetic part we have (\ref{Lambda3}).

\subsection{Stability analysis for unaligned systems $\bbar=0$}
\label{app.SAUS} We have proposed that, given a sufficient gap
between $\lambda^F_+$ and $\lambda^F_-$, then one alignment of the
ferromagnetic solution may be considered dominant and the absolute
value and fluctuations in the converse direction may be ignored.
However, if these values are comparable one must consider an
expansion in an additional order parameters to understand the
ferromagnetic phase: $\tal \rightarrow \{\qbaral,\qal\}$. To
correct the auxiliary system we make the substitution of the type
$\tal^i=\langle (\qal + b \qbaral)^i \rangle_b$ for $i^{th}$ order
terms in the entropic part (\ref{Lambdaapp}). There are only two
additional terms at $4^{th}$ and $5^{th}$ order in the free energy to consider, which
couple the two parameters. This is a complicated expression to
evaluate.

Under RS we can anticipate a set of results including
$\{\qal\neq 0,\qbaral=0\}$ and $\{\qal= 0,\qbaral \neq 0\}$ to
dominate based on numerical findings and which are both locally stable (section ~\ref{ssec:UFC}). 
One simple observation is that in the expansion we find that at leading orders, excluding critical points, there is symmetry about the line $\beta J_0 = \rho \mT_1$. By expression of this equality in the $\gamma,\beta$ plane (\ref{app.TP}) we can observe this is skewed (depending on the scaling, e.g. only at subleading order (\ref{Hamgam})) towards higher $\gamma$ with decreasing temperature. If the alignment changes monotonically with $\gamma$ then the point immediately below $\gamma_c$ will be unbiased in alignment at leading order, but favour a dense alignment when the non-linear dependence is included. This qualitative argument appears consistent with results, but the discontinuous nature of the transition requires a consideration of more complicated effects.

\section{Triple point analysis}
\label{app.TP} To discriminate between the effects of different
subsystems one must make an expansion of $\Delta_i$ as a function 
of $\delta\beta$ and $\delta \gamma$ about the triple-point. If we
take $\Delta\gamma = \mu \Delta T$ we can attain an equation for
the critical line (\ref{criticalline}) dependent on the ensemble
details
\begin{eqnarray}
 \Delta_1 &=& \frac{\partial \lambda^F}{\partial \gamma} (\mu \delta \beta)
 + \frac{\partial \lambda^F}{\partial \beta} \delta \beta + O(\delta \beta^2) \\
 \Delta_2 &=& \frac{\partial \lambda^S}{\partial \gamma} (\mu \delta \beta)
 + \frac{\partial \lambda^S}{\partial \beta} \delta \beta + O(\delta \beta^2)\;.
\end{eqnarray}
Which gives an equation for the value in which ferromagnetic instability emerges
\begin{equation}
\begin{array}{cccc}
 \mu &= \frac{-1}{\beta} \left.\frac{\beta \Jo -\beta^2 \Jt + \rho\<\beta x
 (1 - \tanh(\beta x) +\tanh^2(\beta x) -\tanh^3(\beta x) )\>_{\phi(x)}}{ -\beta
\Jo'
 + \beta^2 \Jt'/2 + \rho\<\beta x (1 - \tanh(\beta x) +\tanh^2(\beta x)
 -\tanh^3(\beta x)) \>_{\phi'(x)}}\right|_{\beta_C,\gamma_C} \ .
\end{array}
\end{equation}
The derivatives in the denominator in $\Jo,\Jt$, and distribution
$\phi$, denoted by $'$ are with respect to $\gamma$. These are
taken straightforwardly with linear scaling of couplins (\ref{Hamgam}). The
expression is quite complicated to interpret, but may be simplified 
further using the criteria for triple point existence
(\ref{1spinorder}) and (\ref{2spinorder}). These expression may be
used to find a triple point for a particular system in variation of 
some parameters ($\gamma,\beta$) and we can then test whether the dense or spin induced order
dominates as temperature is lowered.

\begin{figure*}[htb]
\begin{center}
\includegraphics*[width=0.6\linewidth]{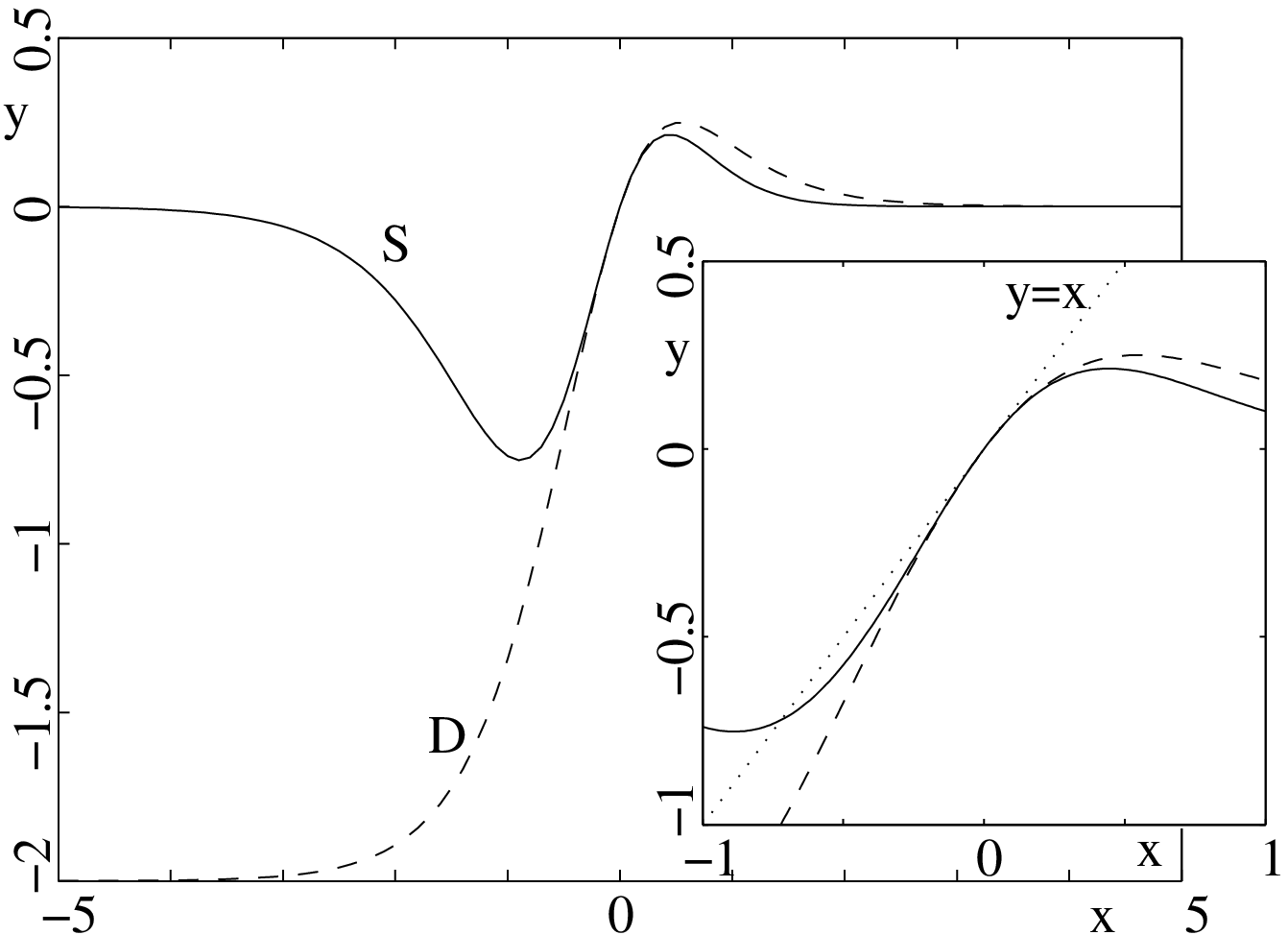}
 \caption{\label{SANDDfig} The dashed lines indicate the quantity $D$
(\ref{Djudge}) and the solid lines the quantity $S$
(\ref{Sjudge}). If one has a particular distribution of sparse
couplings $\phi$ then at the triple-point one can integrate over
these distributions to determine the quantities and hence the
nature of reentrant behaviour -- whether the weakly ordered
systems are dominated by dense or sparse type order, we find the former in canonical cases. If the
variance of the distributions is sufficient the integral for both
quantities is negative. If the distribution is of positive mean
with low variance both quantities will be positive. It is possible
to generate many distributions for which the above generalisations
do not hold, and reentrant behaviour of an alternative type may
occur.}
\end{center}
\end{figure*}

 Rewriting the transition line for linear scaling of the Hamiltonian
(\ref{Hamgam}), and resubsituting (\ref{1spinorder}) and
(\ref{2spinorder}) we find for the aligned system ($\bbar=1$):
\begin{eqnarray}
 (\gamma-\gamma_C) &=& \mu (\beta-\beta_C) \\
 \mu &=& \frac{1}{\beta} \left.\frac{ - \<D\> + \<S\> }{\<D\>/(1-\gamma) +
\<S\>/\gamma}\right|_{\beta_C,\gamma_C} \\
 D &=& \tanh(\beta \gamma x) - \tanh^2(\beta \gamma x) \label{Djudge}\\
 S &=& \beta \gamma x (1 - \tanh(\beta \gamma x) +\tanh^2(\beta \gamma x)
-\tanh^3(\beta \gamma x) )\label{Sjudge}
\end{eqnarray}
Where the distribution $\phi(x)$, over which averages occur, is by
the decomposition (\ref{Hamgam}) now a distribution independent of
$\gamma$. Interestingly, this value is independent of the
connectivity $\rho$, thus percolation properties appear not to be
an important effect in the RS order stability in the vicinity of
the triple-point. This connectivity independence is true for more
general scaling scenarios than~(\ref{Hamgam}); whenever the
derivatives of $\Jo$,$\Jt$ with respect to $\gamma$ are
proportional to $\Jo$,$\Jt$, one can make substitutions to
write the expression as functions of only $\phi,\gamma,\beta$. For $\bbar \neq 1$, 
this is not necessarily possible though certain properties are similar.
The expressions $S$ and $D$ are shown in figure~(\ref{SANDDfig}),
it is clear that if $\phi$ is negative in mean or has large
variance then for most distributions $\<D\>$ and $\<S\>$ are
negative, with $\<D\> < \<S\>$. If the distribution is positive in
mean and of relatively small variance then $\<D\>$ and $\<S\>$
will be positive, with $\<D\>$ the larger. However, this is not a
general result and exceptions may be constructed.

\section{Population dynamics numerical implementation}
\label{app.PD}
The population dynamics is implemented such that the fields are
assigned randomly at time $t=0$ in three histograms (real replica) simultaneously.
 We consider the development from
initial conditions corresponding to frozen ferromagnetic, SG and
near paramagnetic (high temperature) initial condition histograms. 
The corresponding Gaussian distributions are
\begin{eqnarray}
 P(h^{(0)}_i) &=& {\cal N}(1000, 0) \qquad , \qquad {\cal N}(0, 1)
\qquad \mbox{and} \qquad {\cal N}(0, 0.001) \; ,
\end{eqnarray}
We also add an
additional (fourth) histogram with ferromagnetic
order along the dense alignment when $\bbar\neq1$,
and initialise $b_i^{(0)}$ to
$\pm 1$ with probabilities determined by $\bbar$.

Since the representative low-temperature order parameters are
overestimated in the former two, and underestimated in the latter,
we can be confident a solution converged to by all three will not
be systematically biased. However, in the case of multiple locally stable solutions 
these replica remain disparate, but cover a representative set of solutions quite well (e.g. $\bbar=0$).
This parallelism is at the cost of runtime, but we found
drift, or metastability, to be prevalent effects justifying the method. 

It is also useful to consider
fluctuations of the solutions within a single histogram in successive
time steps ($N$ field updates). We judge convergence in both cases through consideration of the order parameters $q_1$, $\qbaro$, and $q_2$. This is
insufficient for a system involving sparse interactions since the distribution is then not Gaussian, but
appears nevertheless to be quite robust and comparable to other heuristic methods attempted. The criteria for
termination of the algorithm is that the distributions based on
all 3 (4) initial conditions and between successive updates converges in
$q_1$,$\qbaro$ and $q_2$ to within a worst case absolute precision of $0.01$.
Finite size fluctuations and runtime constraints prevent a signigicantly stronger precision.

The field updates of (\ref{fieldupdateline1}) are random and non-sequential
 -- but ensuring that in each time step, each field is
updated exactly once.

We found in almost all cases that histograms converged to a
single unique distribution (upto finite size errors), and that this convergence was robust. When convergence criteria is met we analyse the histogram (replica) of lowest free energy as representative. Measurables are determined
following convergence based on 20 samples over 20 time steps. For certain systems we ran upto $20$ independent runs (different pseudo random update sequences) to check the properties were unique and fully explore the space.

In the case of non-convergence by the moments criteria we halt the algorithm
after $500$ time steps and consider the histogram that corresponds to the lowest
free energy (amongst the different initial condition histograms),
 calculating quantities based on this choice. In the case that metastable
solutions exists this is sufficient to identify the histogram corresponding to a dominant state. This occured only
for certain systems with opposite ferromagnetic alignment ($\bbar=0$). In
all other cases it appeared solutions are in the vicinity of the correct
saddlepoint but converging with different bises; thus identifying the solution of lowest free energy
does not guarantee improved measurement of the statistics in this case.
However, we judge the statistics collected to be close in most cases.

\subsection{Stability} We associate with each field in the population dynamics a
perturbation $\delta h_i^2$, these are initially chosen with a Gaussian
distribution independent of the field. With $\gamma>0$ convergence to a non-Gaussian joint distribution of
perturbations and fields must be considered, we observed this to occur very quickly as can be determined in
correlation functions and kurtosis, e.g.
$\<\delta (h_i^{t})^2 (h_i^{t})^2\>_C$ and $\<\delta h_i^4\>$, but could not produce a robust measure of convergence of the fluctuation distribution.
Following convergence in the order parameters, we introduce the
field and following $10$ time steps collect data over 20 time steps.
Distribution of perturbations are carefully renormalised throughout analysis to account
for the non-sequential updates, and the renormalisation after each time-step
provides an estimate for $\chi_2$. Certain other numerical
hacks must be brought to bear to prevent divergences in the cases where $\chi_2$
is not close to $0$.
\clearpage

\section*{References}
\bibliographystyle{mybib}
\bibliography{SD}
\end{document}